\documentclass[fleqn,usenatbib]{mnras}

\usepackage{newtxtext,newtxmath}

\usepackage[T1]{fontenc}

\DeclareRobustCommand{\VAN}[3]{#2}
\let\VANthebibliography\thebibliography
\def\thebibliography{\DeclareRobustCommand{\VAN}[3]{##3}\VANthebibliography}

\usepackage{graphicx}	
\usepackage{amsmath}	
\usepackage{pgffor}
\usepackage{xtab,booktabs}
\usepackage{longtable}

\newcommand{\DsDl}{$\mathcal{D}_{S}/\mathcal{D}_{L}\,$}
\newcommand{\DtoZ}{$\mathcal{D}/Z\,$}
\newcommand{\Dtot}{$\mathcal{D}_{\text{tot}}\,$}
\newcommand{\NoDiff}{\emph{NoDiff}}
\newcommand{\WeakDiff}{\emph{Diffx0.1}}
\newcommand{\Diff}{\emph{Diffx1}}
\newcommand{\StrongDiff}{\emph{Diffx10}}
\newcommand{\HiRes}{\emph{HiRes}}

\defcitealias{1977ApJ...217..425M}{MRN}
\defcitealias{2017MNRAS.466..105A}{A17}
\defcitealias{2019MNRAS.482.2555H}{HA19}
\defcitealias{2020MNRAS.491.3844A}{AHN20}

\title[Dust Diffusion]{Dust diffusion in SPH simulations of an isolated galaxy}

\author[L. Romano et al.]{
Leonard E. C. Romano$^{1,2}$\thanks{E-mail: leonard.romano@tum.de},
Kentaro Nagamine$^{2,3,4}$, 
Hiroyuki Hirashita$^{5}$
\\
$^{1}$Physik-Department, Technische Universität München, James-Franck-Straße, 85748 Garching, Germany\\
$^{2}$Theoretical Astrophysics, Department of Earth and Space Science, Osaka University, 1-1 Machikaneyama, Toyonaka, Osaka 560-0043, Japan\\
$^{3}$Kavli IPMU (WPI), The University of Tokyo, 5-1-5 Kashiwanoha, Kashiwa, Chiba 277-8583, Japan\\
$^{4}$Department of Physics and Astronomy, University of Nevada, Las Vegas, 4505 S. Maryland Pkwy, Las Vegas, NV 89154-4002, USA\\
$^{5}$Institute of Astronomy and Astrophysics, Academia Sinica, Astronomy-Mathematics Building, AS/NTU, No. 1, Section 4, Roosevelt Road, Taipei 10617, Taiwan \\
}

\date{Accepted XXX. Received YYY; in original form ZZZ}

\pubyear{2021}

\begin{document}
\label{firstpage}
\pagerange{\pageref{firstpage}--\pageref{lastpage}}
\maketitle

\begin{abstract}
We compute the evolution of the grain size distribution (GSD) in a suite of numerical simulations of an isolated Milky-Way-like galaxy using the $N$-body/smoothed-particle-hydrodynamics code {\sc Gadget4-Osaka}. The full GSD is sampled on a logarithmically spaced grid with 30 bins, and its evolution is calculated self-consistently with the hydrodynamical and chemical evolution of the galaxy using a state-of-the-art star formation and feedback model. In previous versions of this model, the GSD tended to be slightly biased towards larger grains and the extinction curve had a tendency to be flatter than the observations.
This work addresses these issues by considering the diffusion of dust and metals through turbulence on subgrid scales and introducing a multi-phase subgrid model that enables a smoother transition from diffuse to dense gas. We show that diffusion can significantly enhance the production of small grains and improve the agreement with the observed dust extinction curve in the Milky Way.

\end{abstract}




\section{Introduction}

The distribution of dust in the interstellar medium (ISM) is coupled to the evolution of galaxies. Even though only about one percent of the gas in the ISM is condensed in the form of dust grains, they play an enourmous role for galaxy evolution. There is a plethora of processes occuring on the surfaces of dust grains, which greatly affect the chemical and thermal history of galaxies. One of these processes, the efficient formation of molecular hydrogen on grain surfaces has important consequences for star formation as it leads to the formation of molecular clouds which eventually form stars \citep{2002MNRAS.337..921H, 2004ApJ...604..222C, 2011ApJ...735...44Y}. By absorbing ultra-violet (UV) light and reemitting in the infrared (IR), the dust also acts as a coolant itself, leading to star formation in dusty clouds \citep{2005ApJ...626..627O}, shaping the initial mass function (IMF) \citep{2012MNRAS.419.1642C} and spectral energy density (SED) of observations of galaxies \citep{2005MNRAS.362..592T} as well as determining the typical range of stellar masses \citep{2006MNRAS.369.1437S}. Since smaller grains have a larger surface area per unit mass than larger grains, the total available surface area for these processes is not just determined by the total abundance of dust, but also by the grain size distribution (GSD). In particular, the shape of the GSD affects the efficiency of the rate at which processes occuring on grain surfaces happen \citep[e.g.][]{2011ApJ...735...44Y, 2017ApJ...837...78H} as well as shape of the dust extinction curve \citep[e.g.][henceforth MRN]{1977ApJ...217..425M}. A complete understanding of the evolution of the GSD is thus of critical importance for a complete picture of galaxy evolution.

\citet{2013MNRAS.432..637A} have developed a comprehensive model for the evolution of the GSD in the ISM, in which large dust grains with sizes $a \sim 0.1 \mu\text{m}$ \citep{2012ApJ...745..159Y} condense from the metals in supernova (SN) ejecta \citep[e.g.][]{1989ApJ...344..325K, 2007MNRAS.378..973B, 2007ApJ...666..955N} and the stellar winds of asymptotic giant branch (AGB) stars \citep[e.g.][]{2006A&A...447..553F, 2014MNRAS.439..977V, 2017MNRAS.467.4431D}. These grains are then processed through a number of processes. In sufficiently warm and diffuse gas, large grains are efficiently shattered, leaving behind small fragments \citep[e.g.][]{2009MNRAS.394.1061H}. In the cold, dense and metal enriched ISM small grains can efficiently grow through accretion of gas phase metals and stick together to form larger grains in a process known as coagulation \citep[e.g.][]{2012MNRAS.422.1263H}. 
In regions where the metallicity exceeds $\sim 0.1\, Z_{\sun}$, accretion can lead to a rapid growth of small grains until 
it saturates once most of the readily depleted gas phase metals have been used up \citep[e.g.][]{1998ApJ...501..643D, 2008A&A...479..453Z, 2012MNRAS.424L..34K}, increasing the abundance of small grains. Subsequent coagulation shapes the GSD by populating the intermediate size range, smoothing it towards a power-law shape similar to the \citetalias{1977ApJ...217..425M} grain size distribution. In the hot gas of the circum-galactic medium (CGM) or gas heated by SN shocks, grains can be evaporated through sputtering processes \citep[e.g.][]{1994ApJ...431..321T}, while in star-forming regions astration -- the subsumption of dust into the interior of stars -- reduces the dust abundance.

The relative efficiencies of the processes listed above strongly depend on the local physical conditions, like temperature, density and metallicity. Therefore in order to understand the spatial variation of the GSD and the effect of dust transport to different environments, the evolution of the dust needs to computed consistently with the hydrodynamic evolution of the ISM. To this end, hydrodynamic simulations have been used to study the evolution of the dust alongside the evolution of the ISM in isolated galaxies \citep[e.g.][henceforth HA19 and AHN20]{2019MNRAS.482.2555H, 2020MNRAS.491.3844A} and in cosmological simulations of structure formation \citep[e.g.][]{2018MNRAS.478.4905A, 2019MNRAS.485.1727H}.
A variety of methods have been used by many researchers to model the dust.  \citet{2015MNRAS.451..418Y} used cosmological zoom-in simulations to study the dust distribution in high-redshift galaxies, assuming a constant dust-to-metal ratio. \citet{2015MNRAS.449.1625B} treated dust as a separate massive particle species, which is affected by a drag force due to gas and radiation pressure due to stars. The dust particles in their model evolve through dust growth and destruction, but also affect star formation and feedback. \citet{2016MNRAS.457.3775M} treated dust as a component dynamically coupled to the gas in their cosmological zoom-in simulations. They considered growth and destruction processes, finding that in the absence of growth processes the dust mass is heavily suppressed. Furthermore, \citet{2017MNRAS.468.1505M} studied the statistical properties of dust in a large sample of galaxies from a suite of cosmological simulations. They were able to reproduce the present-day dust abundance in galaxies, but tend to underestimate the dust mass at high redshift. \citet{2018MNRAS.478.2851M} combine previous efforts in modelling the evolution of dust using `live' dust particles, each of which samples a local GSD. They performed promising test calculations of an isolated galaxy, but further implementation of e.g. stellar feedback, is needed to provide results to be compared to observations.
\citet{2021MNRAS.501.1336H} post-processed a cosmological simulation with a model of the GSD to study the extinction law in Milky-Way-like galaxies. \citet{2021MNRAS.507..548L} performed a similar study by directly implententing the evolution of GSD in a cosmological zoom-in simulation.
Both of these studies find a large diversity of extinction laws, with bump strengths and UV slopes that are comparable to observations in the Milky Way.

\citet[][hereafter A17]{2017MNRAS.466..105A} studied the evolution of small and large dust grains in a simulation of an isolated galaxy. Their model is based on the \textit{two-size approximation} \citep{2015MNRAS.447.2937H}, in which the GSD is sampled by two bins referred to as `small' and `large' grains to save the computational cost. Their radial dust profiles are in agreement with observations of late type galaxies by \citet{2012MNRAS.423...38M} and their results predict a variation of the GSD with the galactic radius. \citet{2017MNRAS.469..870H} followed up on these results, but evolved carbonaceous and silicate dust separately, studying the dynamical evolution of extinction curves in the same isolated galaxy. They find a steepening of the extinction curve at intermediate age and metallicity, at which the dust is efficiently processed by shattering and accretion. \citet{2018MNRAS.474.1545C} post-processed the simulation by \citetalias{2017MNRAS.466..105A} to compute the equilibrium abundances of H$_2$ and CO, finding that H$_2$ fails to trace the star formation rate (SFR) at low metallicity because under such conditions H$_2$ is confined to dense, compact clouds. The two-size approximation has also been applied in cosmological simulations \citep[e.g.][]{2018MNRAS.478.4905A, 2018MNRAS.479.2588G, 2019MNRAS.484.1852A, 2019MNRAS.485.1727H, 2021MNRAS.503..511G}. These simulations largely explained a number of observed relations, like the relation between the dust-to-gas ration and metallicity, the dust mass function and the evolution of the comoving dust mass density.

\citet{2019MNRAS.485.1727H}, adopting the same approximation in a cosmological simulation, also attempted to investigate extinction curves. However, the limited freedom in the GSD did not allow them to predict detailed extinction curve shapes. Recently, \citetalias{2020MNRAS.491.3844A} adressed this, by simulating the evolution of the full GSD sampled at 32 logarithmically spaced grid points in a simulation of an isolated galaxy. They showed that some of the results, like the evolution of the dust-to-gas ratio with metallicity and radial profiles of the dust-to-gas ratio and dust-to-metal ratio are unaffected by the two-size approximation. The main focus of their work was laid on the spatial and temporal evolution of the grain size distribution. They find that dust evolution happens in three stages dominated by stellar yield, accretion and coagulation, respectively. They also studied the evolution of extinction curves, in the dense and diffuse medium and find that the extinction curve in the dense medium first becomes steeper than in the diffuse medium at intermediate times and then flattens as the GSD settles to the MRN power law while the extinction curve in the diffuse medium steepens.

\citet{2020A&A...636A..18R} compared the results of the simulations from \citet{2017MNRAS.466..105A} and \citet{2017MNRAS.469..870H} to spatially resolved observations of nearby spiral galaxies and the results from \citet{2019MNRAS.485.1727H} to the integrated properties of their sample galaxies. They found that while the simulations tend to agree with the observed total dust abundances at high metallicity, the agreement gets worse at low metallicity. Moreover, they find that there are some discrepancies between the observed and the simulated small-to-large grain ratios, especially in galaxies with high stellar mass. Rela\~no et al. (2022; in prep.) take an unprecedented observational sample of 247 local galaxies from five state-of-the-art galaxy surveys and compare their dust properties obtained by SED fitting to the results of the cosmological simulations of \citet{2018MNRAS.478.4905A} and \citet{2019MNRAS.485.1727H}.
They find that the simulations tend to overestimate the dust mass in the high stellar mass regime. They also find that the small-to-large grain ratios predicted by the simulations are consistent with a subsample of their galaxy sample, which exhibits lower small-to-large grain ratios at high stellar and dust mass. However, another subsample of the galaxies with high small-to-large grain ratios cannot be explained by the simulations. It remains unclear how to explain the large scatter in the relation between the small-to-large grain ratio at high stellar and dust mass. It is encouraging to see that such efforts to reconcile simulations and highly detailed observations are becoming possible, especially in the light of high resolution observations from ALMA and integral field spectroscopy.

While these simulations successfully explain a large variety of observational results, like the evolution of the dust-to-gas ratio with metallicity or the radial dust profiles in nearby galaxies, there are still a number of issues that need to be addressed. \citetalias{2019MNRAS.482.2555H} point out that the production of small grains is insufficient in order to explain the Milky Way extinction curve and that coagulation needs to be more efficient in order to explain the observational trend of flatter extinction curves in denser gas.
While this issue turned out to be less problematic than originally expected \citepalias{2020MNRAS.491.3844A}, the GSD in their simulations still tend to be slightly biased toward large unprocessed grains and thus their median extinction curves are slightly flatter than the observations. 

Given that the models used for dust processing seem to be largely comprehensive, it is worthwhile considering whether or not the results might be biased due to the hydrodynamical scheme. Most groups, studying the evolution of dust in the ISM used Lagrangian hydrodynamics schemes. In these methods, the fluid is mass-discretized and thus, by definition mass mixing between the fluid tracers and therefore also chemical mixing is absent. A popular method to account for the missing advection in Lagrangian codes is the diffusion prescription of \citet{2010MNRAS.407.1581S}, which itself is based on the Smagorinsky-Lilly model \citep{1963MWRv...91...99S}. The
impact of a numerical diffusion prescription in Lagrangian schemes has been investigated by numerous groups \citep[e.g.][]{2018MNRAS.480..800H, 2018MNRAS.474.2194E, 2021ApJ...917...12S}. 
\citet{2018MNRAS.480..800H} find that including metal mixing does not affect any of the gross galactic properties like star formation or gas dynamics but can influence the abundance ratio distributions as discussed in detail by \citet{2018MNRAS.474.2194E}. \citet{2021ApJ...917...12S} compare the metal enrichment of the CGM in simulations of an isolated galaxy, with the mesh-based code ENZO and the particle based codes {\sc Gadget-2} and Gizmo-PSPH. They find that the inclusion of a subgrid model for turbulent diffusion between the particles is required in particle based codes in order to achieve the same level of mixing as in the mesh-based code. Given that in the dust evolution model by \citet{2013MNRAS.432..637A} interstellar processing is treated in the dense and diffuse medium separately,
including mixing between the dense and the diffuse medium would provide a natural channel to accelerate dust processing and enhance the interplay between the processes. In particular, the interplay between shattering in the diffuse ISM and accretion in the dense ISM plays an important role in enhancing the small-grain abundance \citepalias{2020MNRAS.491.3844A}. This
might reduce the bias towards 
unprocessed large grains.

The goal of this study is to address the previously reported issues in the dust model employed by \citetalias{2019MNRAS.482.2555H} and \citetalias{2020MNRAS.491.3844A}. To this end we study, to what extent fluid mixing by turbulent diffusion can affect dust processing.
At the same time, we examine the robustness of \citetalias{2020MNRAS.491.3844A}'s conclusions against the inclusion of diffusion. For example, we test if the different grain size distributions between the dense and diffuse ISM are maintained or not.
Additionally we address the low efficiency of coagulation and accretion reported by \citetalias{2019MNRAS.482.2555H} by recalibrating the subgrid recipe for the modelling of the unresolved dense clouds.
Apart from the issues that these changes are meant to address, we expect to see additional features in the spatial distribution of dust and metals, due to mixing between the galactic disk and the CGM.

This paper is structured as follows. We describe the simulation setup and the adopted physical models in Section~\ref{sec:methods}. In Section~\ref{sec:results}, we present our simulation results. In Section~\ref{sec:Extinction} we compare our results with the observed extinction curve in the Milky Way. In Section~\ref{sec:discussion}, we discuss our results and conclude by summarizing our findings. Throughout this paper, we adopt a value of $Z_{\sun} = 0.01295$ for the solar metallicity consistent with the default value in Grackle-3 \citep{2017MNRAS.466.2217S}. We adopt a value of $0.03\,\micron$ for the grain radius which separates `large' and `small' grains.
  
\section{Methods}
\label{sec:methods}

\subsection{Hydrodynamic Simulation}\label{sec:simulation}
We study the evolution of a simulated isolated galaxy using the {\sc Gadget4-Osaka} smooth particle hydrodynamics (SPH) simulation code, which is based on  a combination of the {\sc Gadget-4} code \citep{2021MNRAS.506.2871S} and the {\sc Gadget3-Osaka} feedback model \citep[]{2017MNRAS.466..105A,2019MNRAS.484.2632S, 2021ApJ...914...66N}. 
We treat the star formation and production of dust and metals self-consistently with the hydrodynamic evolution of the system, accounting for the effects of SN feedback. In our simulations, the relative motion of dust and gas is neglected, and instead it is assumed that the dust is carried along by the gas particles (i.e.\ tight coupling between them). Gas cooling and primordial chemistry are treated using the {\sc Grackle-3} library \footnote{\label{footer:Grackle3} \href{https://grackle.readthedocs.org/}{https://grackle.readthedocs.org/}}\citep{2017MNRAS.466.2217S}, which provides a 12 species non-equilibrium chemistry solver for a network including reactions among the species H, H$^+$, He, He$^+$, He$^{2+}$, e$^-$, H$_2$, H$^-$, H$_2^+$, D, D$^+$ and HD.
Photo-heating, photo-ionization and photo-dissociation due to the UV background radiation at $z = 0$ from \citet{2012ApJ...746..125H} is taken into account. The cooling rate for metal-line cooling is linearly scaled with total metallicity, i.e. we do not distinguish between gas-phase metals and metals condensed into dust grains.

In order to estimate the effect diffusion on the evolution of the GSD, we evolve an isolated Milky Way-like galaxy for $2\,\text{Gyr}$, beyond which the GSD does not evolve much. The initial conditions (ICs) are the low resolution ones of the AGORA collaboration \citep{2016ApJ...833..202K}, but following \citet{2021ApJ...917...12S} we have added a hot ($T = 10^6\, \text{K}$) gaseous halo with a mass of $\sim 10^9 \, \text{M}_\odot$. The gas is initially assumed to have primordial abundances ($X_{\text{H}} = 0.752$, $X_{\text{He}} = 0.248$) and no dust. While this setup is admittedly quite unrealistic, the main goal of this study is to appreciate the effect diffusion has on the evolution of dust.
A zero-metallicity IC, as also adopted by \citetalias{2020MNRAS.491.3844A}, serves as a useful experiment
to cover all important stages of dust enrichment, and provides a way of experimenting the effect of diffusion at various stages of metal and dust enrichment.
A detailed assessment of the dependence on the ICs is given in a companion paper \citep{2022arXiv220205521R}. We let the ICs relax adiabatically for $500 \text{Myr}$, in order to avoid numerical artifacts due to density fluctuations. We then enable the subgrid physics and evolve the relaxed ICs for another $2 \,\text{Gyr}$. We run four simulations with and without turbulent diffusion of metals and dust and one simulation with tenfold mass refinement with diffusion. The gravitational softening length is set to $\epsilon_\text{grav} = 80\, \text{pc}$ in the low resolution runs and to $\epsilon_\text{grav} = 40\, \text{pc}$ in the high resolution run. We do not allow the SPH smoothing length to get smaller than a tenth of the gravitational softening length.

In order to produce similar amounts of stars and metals over the simulated period in all runs, we need to recalibrate the star formation and feedback model parameters in the run with higher resolution. To this end, we increase the threshold density for star formation to occur according to Larson's law \citep{1981MNRAS.194..809L}, i.e. effectively double its value to $n_{\text{H, th}} = 20\,\text{cm}^{-3}\,$
and halve the number of energy injections due to type-II SN feedback per star particle in order to achieve similar energy injections per feedback event. This treatment leads to a similar star- and metal production as discussed below in section \ref{sec:SFhistory}.

\subsection{Dust Processing}\label{sec:dustprocessing}

We are using an updated version of the model used by \citetalias{2019MNRAS.482.2555H} and \citetalias{2020MNRAS.491.3844A} for the evolution of the GSD, which is based on the model by \citet{2013MNRAS.432..637A}. The processes considered are the stellar dust production, shattering in the diffuse ISM, growth by accretion and coagulation in the dense ISM, and destruction in SN shocks and through thermal sputtering. On the scales considered in this work, it is usually safe to assume that the gas and dust are dynamically coupled \citep{2018MNRAS.478.2851M}. We therefore neglect the relative dynamics of dust and gas and simply treat the dust as a property of the gas particles. Another important simplification is that we do not distinguish between different dust species. This makes it difficult to reliably distinguish metals which would readily condense into dust grains like C and Si, from metals which are probably less likely to do so.
This would cause a systematic overestimate of dust-to-metal ratio at late times, when dust growth by accretion is important, but would not change the relative significance of this process at various ages and metallicities.

Just as in previous versions of the model the GSD is expressed in terms of the grain mass distribution $\rho_d\left(m, t\right)$ which is defined such that $\rho_d\left(m, t\right) \text{d}m$ is the mass density of grains with mass $m \in \left[m, m+\text{d}m\right]$. The grains are assumed to be compact and spherical such that $m\left(a\right) = \left(4\pi/3\right) s a^3$, where the bulk density $s = 3.5\, \text{g}\,\text{cm}^{-3}$ appropriate for silicates is adopted \citep{2001ApJ...548..296W}. With this definiton of the grain mass distribution, the dust-to-gas ratio is
\begin{equation}
    \mathcal{D}_{\text{tot}}\left(t\right) = \int \frac{\rho_d\left(m, t\right)}{\rho_{\text{gas}}\left(t\right)} \text{d}m.
\end{equation}
We sample the GSD with 30 logarithmically spaced bins in the range of $3 \times 10^{-4}$ -- $10 \,\micron$ and enforce vanishing boundary conditions through the use of a virtual empty bin at each boundary.

The evolution of the GSD through stellar yield, shattering, coagulation and accretion is modelled in the exact same way as described in \citetalias{2020MNRAS.491.3844A} and we refer the interested reader to their paper. In their model, whenever metals are ejected into the ISM by SN explosions or stellar winds from AGB stars, a fraction $f_\mathrm{in} = 0.1$ of the ejected metals \citep{2011EP&S...63.1027I, 2013MNRAS.436.1238K} are assumed to be condensed into dust grains following an initial log-normal GSD centred around $a_{0} = 0.1\,\micron\,$ and with a variance of $\sigma = 0.47$ \citep{2013MNRAS.432..637A}. In the diffuse ISM large grains are shattered into smaller fragments, while in the cold and dense ISM small grains can grow through accretion of gas phase metals or stick together to form larger grains in a process known as coagulation. Dust is lost due to star formation and strong shocks due to SN feedback.  
We modified the model for estimating the multiphase structure on subgrid scales used to determine the strength of accretion and coagulation, added thermal sputtering and modified the treatment of destruction in SN shocks. These changes are described below.

\subsubsection{Two-phase ISM subgrid model}\label{sec: MultiPhase}
Processes like coagulation and accretion can only happen in a sufficiently cold and dense medium, while other processes like shattering are more efficient in the warm diffuse medium. Since the former presently cannot be resolved in our simulations, a subgrid model needs to be employed in order to resolve the effects associated to such environments.

\citetalias{2020MNRAS.491.3844A} assume that cold ($T < 10^3\, \text{K}$) and dense ($n_\text{H} > 10\, \text{cm}^{-3}$) gas particles host dense clouds with $n_\text{H} = 10^3\, \text{cm}^{-3}$ and $T = 50\, \text{K}$ on unresolved scales, making up $f_\text{cloud} = 0.5$ of the particle's
mass. Analysis of snapshots of their simulations revealed that the global fraction of dense gas rarely exceeds $1 \,\%$. Assuming that this dense gas traces the amount of molecular gas, this puts an upper limit of $\sim 1 \,\%$ on the molecular gas fraction, in strong disagreement with the typical value of $f_{\text{H}_2} \sim 20 \,\%$ for Milky Way-like spiral galaxies \citep{2018MNRAS.476..875C}.

Lowering the density threshold can lead to slightly better agreement, but also leads to undesirably large amounts of accretion and coagulation in relatively diffuse gas $n_\text{H} \leq 1 \,\text{cm}^{-3}$. Instead, if we assume that dense gas traces H$_2$, then previous results by \citet{2011ApJ...728...88G} and \citet{2014ApJ...795...37G} indicate that denser gas hosts more dense clouds. We thus model this trend by assuming that in cool gas ($T < 10^4\, \text{K}$) the dense fraction increases linearly with the density until it saturates at large densities as  
\begin{equation} \label{eq: fdense}
    f_\text{cloud} = \text{min}\left(\alpha\, n_0, 1\right),
\end{equation}
where $n_0 = n_\text{H} \left[\text{cm}^{-3}\right]$, and alpha is a parameter to set the slope. We have found that $\alpha \sim 0.12$ leads to a global dense gas fraction of $\sim 20 \,\%$ at our fiducial resolution. At higher resolution, as more of the gas reaches higher densities, slightly lower values are preferred. In the high resolution run we thus set $\alpha = 0.08$, in order to match the time evolution of the global dense gas fraction to the lower resolution case.

Another detail that \citetalias{2020MNRAS.491.3844A} neglected in their study was that the rest of the gas in a gas particle hosting dense clouds must be warmer and more diffuse than the densities and temperatures obtained from the hydrodynamics, in order to be consistent. In order to see this, consider a gas particle with fixed mass $m$, average number density $n$ and temperature $T$. The volume occupied by the particle is $V \sim m / n$, omitting constant factors. The condition that the union of the diffuse and the dense medium (indicated by subscripts `diff' and `dense', respectively) fill up the volume occupied by the gas particle reads 
\begin{equation}\label{eq:volume}
    V = V_\text{diff} + V_\text{dense}.
\end{equation}
Conservation of internal energy implies 
\begin{equation}\label{eq:energy}
    pV = \left(pV\right)_\text{diff} + \left(pV\right)_\text{dense},
\end{equation}
where $p_i = n_i T_i$ is the pressure in each component. Given that the fraction of the mass present in the diffuse medium is $f_\text{diff} = 1 - f_\text{cloud}$, equation \ref{eq:volume} leads to the relation between the densities in the diffuse and dense medium
\begin{equation}
    \frac{1}{n} = \frac{f_\text{diff}}{n_\text{diff}} + \frac{f_\text{cloud}}{n_\text{dense}},
\end{equation}
which in the typical case that $n_\text{dense} \gg n$ implies $n_\text{diff} \sim f_\text{diff}\, n$. Similarly eq. \ref{eq:energy} implies that the temperature of the gas particle is the mass weighted temperature of the different ISM components
\begin{equation}
    T = f_\text{diff} T_\text{diff} + f_\text{cloud} T_\text{dense}.
\end{equation}
In the typical case, where $T \gg T_{\text{dense}}$, this implies $T_\text{diff} \sim T / f_\text{diff}$.
We adjust the efficiencies of processes happening in either medium by attenuating the corresponding reaction rates with the respective mass fractions. We neglect processes happening in either medium if the mass fraction falls below $1 \,\%$ in order to save computation time. We note that a similar, but more detailed model for the multi-phase nature of star formation and dust physics on subgrid scales has recently been applied by \citet{2021MNRAS.503..511G}.

\subsubsection{Grain destruction}\label{sec:destruction}

In our model we consider two separate channels of grain destruction, destruction by SN shocks and thermal sputtering in hot gas. Both processes keep the number of grains constant and lead to a shedding of grain surface layers that get ejected in the form of gas phase metals. Thus both processes can be modelled with a continuity equation \citepalias{2019MNRAS.482.2555H}
\begin{equation}\label{eq:sputtering}
    \left[\frac{\partial \rho_d\left(m, t\right)}{\partial t}\right]_{\text{sput}} = \frac{\Dot{m}}{m} \rho_d\left(m, t\right) - 
    \frac{\partial}{\partial m }\left[\Dot{m} \rho_d\left(m, t\right)\right],
\end{equation}
where we estimate $\Dot{m} = - m / \tau_{\text{sput}}\left(m\right)$. We integrate the continuity equation (\ref{eq:sputtering}) by applying the same formulation and integration scheme as \citetalias{2019MNRAS.482.2555H} and \citetalias{2020MNRAS.491.3844A}. 
In the employed feedback model, cooling is temporarily turned off for gas particles subject to SN feedback in order to keep them hot. Since this hot phase cannot be properly resolved, we turn off dust processing as well and only take into account the destruction due to the SN shock after the hot phase has ended. This way the destruction can be regarded as an effective treatment of the dust processing happening in the unresolved hot phase. We employ the same estimate of the SN destruction timescale as \citetalias{2020MNRAS.491.3844A}, which is based on the mass sweeping timescale \citep[e.g.][]{1989IAUS..135..431M}.

Thermal sputtering becomes important at temperatures around $T \sim 10^6\, \text{K}$. Since in our simulation diffusion manages to transport dust out of the cold and dense disk and into the diffuse and hot halo, including thermal sputtering is important in order to prevent the overproduction of small grains through efficient shattering in the halo. 
We approximate the thermal sputtering timescale using equation (14) of \citet{1995ApJ...448...84T}:
\begin{equation}
    \tau_\text{sp}\left(m\right) = \frac{1}{3} \tau_{0, \text{sp}} \left(\frac{a}{0.1 \mu \text{m}}\right) n_0^{-1} \left[ 1 + \left(\frac{T_\text{sput}}{T_\text{gas}}\right)^{\omega}\right],
\end{equation}
where $\tau_{0, \text{sp}} = 9.9 \times 10^4\, \text{yr}$, $T_\text{sput} = 2 \times 10^6\, \text{K}$, and $\omega = 2.5$. Given that this timescale can get much shorter than the dynamical timescale, explicit integration can become very expensive and slow down the simulation. Fortunately the discrete version of the continuity equation provided in the appendix of \citetalias{2019MNRAS.482.2555H} admits an analytical solution that can be used to efficiently integrate the thermal sputtering over long dynamical timesteps. We refer the interested reader to the Appendix A for details about the derivation of this analytical solution.

\subsection{Turbulent diffusion}\label{sec:diffusion}

In Lagrangian simulations by default there are no mass fluxes in between the fluid tracer particles; therefore fluid mixing between particles is generally suppressed leading to discontinuities in quantities like element abundances.
A commonly used computationally rather inexpensive method for smoothing out passive scalar fields like metallicity is to smooth them within a kernel radius similar to the density field \citep[e.g.][]{2005MNRAS.363.1299O, 2019MNRAS.484.2632S}. However, this method fails to capture the transport of abundances beyond the kernel radius and cannot be used in a satisfactory way for quantities that evolve dynamically due to chemistry or processing in the ISM like dust or molecules. Indeed, \citet{2021ApJ...917...12S} have found that explicit inter-particle diffusion of metals due to turbulent mixing is essential for rendering the metal poor gas in the CGM. The importance of diffusion for matching the observed scatter in metal element abundances is widely recognized and many groups working on metal transport with particle based codes have devised ways of modelling diffusion between particles \citep[see e.g.][]{2016A&A...588A..21R, 2018MNRAS.474.2194E, 2021MNRAS.506.4374D}.
In this work, we adopt a turbulent metal and dust diffusion scheme similar to the one used in \citet{2018MNRAS.474.2194E} and \citet{2021ApJ...917...12S} which is based on the Smagorinsky--Lilly model \citep{1963MWRv...91...99S, 2010MNRAS.407.1581S}. In this model, the effect of subgrid diffusion is triggered by the resolved shear between particles. The diffusion operator and the diffusivity are given by
\begin{align}
    \left.\frac{d}{dt}\right|_\text{diff} &= \frac{1}{\rho} \nabla\cdot \left(D \nabla \right)~,\\
    D &= C_{d} \left\|S_{ab}\right\| h^2 \rho~, 
\end{align}
where $C_d$ is a constant parameter setting the diffusion scale $L_{\text{diff}} = \sqrt{C_d} h$, $\left\|\cdot\right\|$ is the Frobenius-norm, $h$ is the SPH smoothing length and $S_{ab}$ is the symmetric, trace-free shear tensor
\begin{equation}
   S_{ab} = \frac{1}{2} \left(\frac{\partial v_a}{\partial x_b} + \frac{\partial v_b}{\partial x_a}\right) - \frac{1}{3} \left(\nabla\cdot \Vec{v}\right) \delta_{ab}.
\end{equation}
Here indices $a, b$ refer to cartesian directions and $v_a$ is the velocity field. We compute $\left\|S_{ab}\right\|$ using the higher-order estimate of the shear tensor computed by {\sc Gadget-4} using matrix inversion \citep{2014MNRAS.443.1173H}. 
We discretize the diffusion equation for the scalar field $A$ following \citet{2004MNRAS.351..423J}
\begin{equation}
    \frac{dA_i}{dt} = \sum_j \frac{m_j}{\rho_i \rho_j} \frac{D_{ij} \left(A_j - A_i\right)}{\left\|\Vec{x}_{ij}\right\|^2}\, \Vec{x}_{ij}\cdot \nabla_i W_{ij},
\end{equation}
where the sum runs over all SPH neighbors, $i$ and $j$ are particle indices, $W_{ij} = 1/2 \left[W\left(\Vec{x}_{ij}; h_i\right) + W\left(\Vec{x}_{ij}; h_j\right) \right]$ is a symmetrised version of the SPH kernel function and $\Vec{x}_{ij} = \Vec{x}_i - \Vec{x}_j$. We make the replacement $\left(D_{i} + D_j\right) \mapsto D_{ij} = \frac{4 D_i D_j}{D_i + D_j}$ following \citet{1999JCoPh.148..227C}, who show that this ensures continuity in the flux at boundaries by effectively selecting the minimum of the two fluxes.

The timescale for diffusion can be estimated from the amount of shear and the degree of mixing between two adjacent fluid elements as
\begin{equation}
    \tau_{\text{diff}, A} \sim 1 \left(\frac{A}{\Delta A}\right) \left(\frac{C_d}{0.01}\right)^{-1} \left(\frac{\left\|S_{ab}\right\|}{100 \,\text{km}\, \text{s}^{-1}\, \text{kpc}^{-1}}\right)^{-1} \text{Gyr},
\end{equation}
which is long compared to the dynamical timescale for well-mixed fluids, but can get extremely short if concentrations vary by many orders of magnitudes. Here $\Delta A$ refers to the difference in the quantity $A$ between the two neighboring fluid elements. In practice this also means that na\"ively requiring a Courant-like timestep criterion
\begin{equation} \label{eq:timestep}
    dt \leq \alpha \frac{A}{\left|\text{d}A/\text{d}t\right|}
\end{equation}
can potentially lead to extremely short timesteps that would unnecessarily slow down the simulation. We thus limit diffusive outflows from low concentration gas with abundances falling below $A_\text{low} = 10^{-4}\, A_\odot$ where $A_\odot$ refers to solar abundance in order to avoid negative abundances. We do not limit the timestep for inflows into low concentration gas. 
In gas with concentrations higher than $A_\text{low}$ we employ a timestep criterion of the form given in equation (\ref{eq:timestep}) with $\alpha = 5$ for inflows ($\text{d}A/\text{d}t > 0$) and $\alpha = 0.5$ for outflows ($\text{d}A/\text{d}t < 0$).
Given the large range of metallicities, this treatment ensures that diffusion does not significantly overshoot, while at the same time avoiding computational overhead. Inflows tend to occur in particles with very low concentrations, where diffusion time-steps are short even with this relatively loose time-step criterion. In principle if our time-step criterion is too loose, overshooting could lead to spurious metal production or destruction. We have verified that the levels of spurious metal production or destruction are negligible, by comparing to a simulation with a more stringent time-step criterion ($\alpha = 0.1$ for both in- and outflows). 

Finally it should be noted that the model is resolution dependent through the definition of the diffusion length scale with respect to the SPH smoothing length which itself scales with the mass resolution. If one requires the diffusion length to be a physical length-scale independent of the resolution, the diffusion parameter $C_d$ has to scale as
\begin{equation} \label{eq: resolution}
    C_\mathrm{d}\propto m_\mathrm{SPH}^{-2/3},
\end{equation}
i.e. $C_d$ needs to be increased as the resolution is refined. A wide range of different diffusion coefficients has been used in the literature \citep[][]{2010MNRAS.407.1581S, 2018MNRAS.474.2194E}. Therefore we estimate the impact of diffusion for a range of values spanning two orders of magnitudes. The full suite of simulations is described in Table~\ref{tab:runs}.

\begin{table}
\caption{List of different simulations}
\label{tab:runs}
\begin{tabular}{lccc}
\hline
Run name & $C_d$ & Resolution & Notes \\
\hline
NoDiff & -- & Low & No diffusion \\
Diffx0.1 & 0.002 & Low & Weak diffusion \\
Diffx1 & 0.02 & Low & Intermediate diffusion \\
Diffx10 & 0.2 & Low & Strong diffusion \\
HiRes & 0.08 & High & High resolution with diffusion \\
\hline
\end{tabular}
\end{table}

\subsection{Particle selection} \label{sec:Selection}
\begin{figure}
\includegraphics[width=0.45\textwidth]{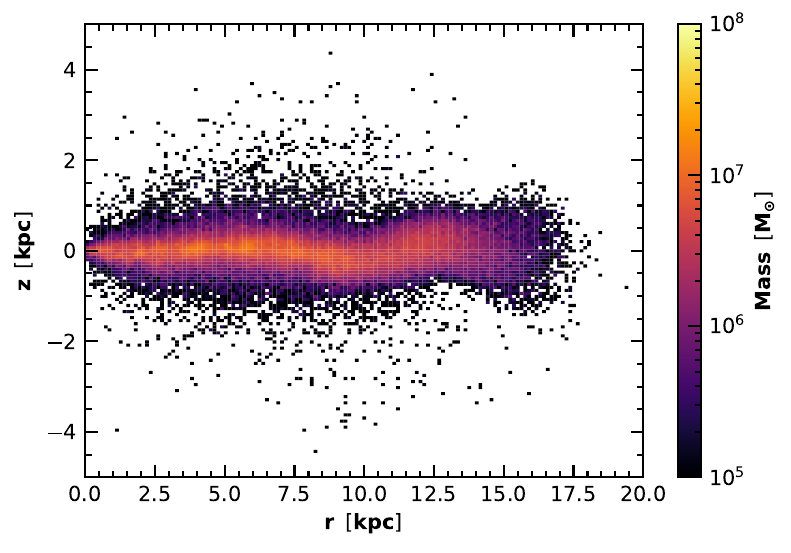}
\caption{The spatial distribution of gas particles belonging to the disk on the $r$--$z$ plane. The color coding indicates the total mass of the particles belonging to each bin. We use the criterion in eq. (\ref{eq:selection}) to select the disk gas particles.}\label{fig:selection}
\end{figure}
The physical properties of the gas within the galactic disk and the halo are very different. As a result, relations between physical quantities like metallicity and dust abundance can be very different within the disk and the halo. While we are mostly interested in the relations in the disk, which makes up most of the gas mass, reliably selecting only particles belonging to this component is not straightforward. While selecting particles through geometric criteria is fairly straightforward, this method is prone to pollution from halo particles in the vicinity of the disk. Another commonly used criterion is based on the orbital circularity parameter $\epsilon = j_z/j_{\text{circ}}\left(E\right)$ \citep{2003ApJ...597...21A}, which is however specific to the {\em thin} disk component and as we show in Appendix B misses a large fraction of the very cold gas in the central part of the galaxy. A method that has proven to be more reliable for our purposes is selection of particles based on the equation of state. The idea is that particles which are part of the disk tend to be denser and colder than particles in the halo. The halo and disk equations of state are well separated in $T$--$n$ space, and thus a rough criterion like
\begin{equation}\label{eq:selection}
    T_\text{gas} < 10^4 \left(\frac{n_\text{H}}{3 \times 10^{-4}\,\text{cm}^{-3}}\right)\,\text{K} 
\end{equation}
can reliably select only particles belonging to the disk as is shown in Figure~\ref{fig:selection}, which shows the distribution of particle coordinates in the $r$-$z$-plane (integrated over the azimuthal angle $\phi$). Here $r$ and $z$ refer to the cylindrical radius as measured from the density weighted centre-of-mas of the gas and the vertical displacement from the disk-plane, respectively.

\subsection{Extinction Curves}

The wavelength dependence of the optical depth of dust is usually expressed in terms of the extinction curve. Extinction curves can be derived from observations and are sensitive to the shape of the GSD, making them a useful tool for relating observations to simulations and constraining the GSD \citep[e.g.][]{2001ApJ...548..296W}. We calculate extinction curves in the same way as \citetalias{2020MNRAS.491.3844A} using the GSD $n_{i}\left(a\right)$, where $i$ indicates the composition of the grains using the same fixed mass ratio of silicate to graphite (54:46), corresponding to the value in the Milky Way. We write the extinction at wavelength $\lambda$ as 
\begin{equation}
    A_{\lambda} \propto \sum_i \int n_i\left(a\right) \pi a^2 Q_{\text{ext}, i}\left(a, \lambda\right) da,
\end{equation}
where $Q_{\text{ext}, i}\left(a, \lambda\right)$ is the ratio of the extinction cross section $\sigma_{\text{ext}, i}\left(a, \lambda\right)$ and the geometrical cross section $\sigma_{\text{g}} = \pi a^2$, which \citet{2001ApJ...548..296W} have evaluated for silicates and carbonaceous grains using Mie theory. 
Their results are tabulated and made publicly available\footnote{\label{footer:ExtinctionParameters} \href{https://www.astro.princeton.edu/~draine/dust/dust.diel.html}{https://www.astro.princeton.edu/~draine/dust/dust.diel.html}}. We normalize $A_\lambda$ to the value in the $V$ band ($\lambda^{-1} = 1.8 \,\micron^{-1}$) in order to cancel out the proportionality constant. We assume that both grain species follow the same GSD following the approach of \citetalias{2020MNRAS.491.3844A}.

\section{Results}\label{sec:results}

To explore the effects of turbulent diffusion, we study both the global assembly history of the dust and metal components and their spatial distribution. We then check how our simulations compare to available observations of dust extinction and small-to-large grain ratios.

\subsection{Star Formation and Metal Enrichment}\label{sec:SFhistory}
\begin{figure}
\includegraphics[width=0.45\textwidth]{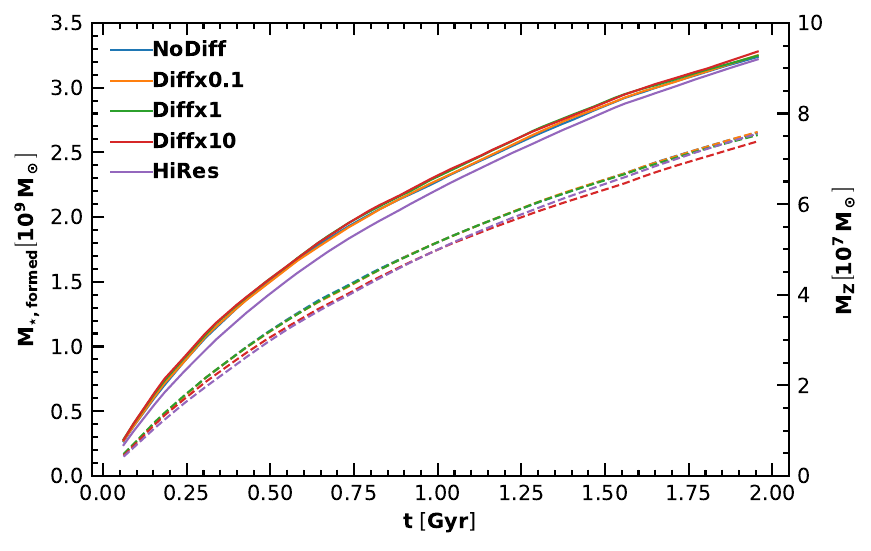}
\caption{The formed stellar mass (solid lines) and the formed metal mass $M_{Z}$ (dashed lines) over the simulated timespan for the different models. Different lines present different models as indicated in the legend (also in figures below). All runs produce similar amounts of stellar and metal mass.}\label{fig:Mstar}
\end{figure}

Metal production in galaxies is linked to the star formation as metals are produced in stars and subsequently injected into the ISM. Thus we require that all runs exhibit a similar star formation history, in order to ensure that differences in the metal distribution are due to differences in the models and not due to different star formation histories. 
The solid lines in Figure~\ref{fig:Mstar} show the time evolution of the formed stellar mass. All runs exhibit very similar star formation histories enabling us to compare the metal distributions among the runs. Despite the similar star formation histories, the metal production depiced by dashed lines in Figure~\ref{fig:Mstar} is slightly lower in the runs with stronger diffusion. In order to understand this difference, it is good to know that the metallicity of the stellar ejecta is almost independent of the stellar metallicity, which implies that metal-poor stars overall introduce more newly formed metals into the ISM than metal-rich stars. 
Thus, differences in the net metal production are likely due to differences in the distribution of stellar metallicities. With stronger diffusion one expects a narrower metallicity distribution with less metal poor stars and thus slightly lower metal production. 

\begin{figure}
\includegraphics[width=0.45\textwidth]{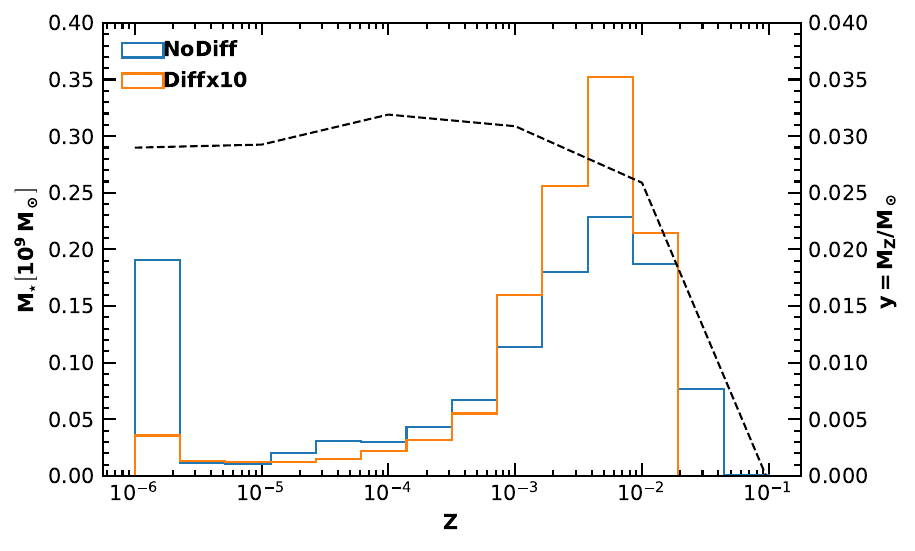}
\caption{The distribution of stellar mass with respect to metallicity at $t = 500 \,\text{Myr}\,$ (solid) in the run without diffusion (blue) and the run with strong diffusion (orange) as well as the metal yield per solar mass as a function of metallicity (dashed).}\label{fig:Zstardist}
\end{figure}

This trend is reflected in Figure~\ref{fig:Zstardist}, which depicts the distribution of stellar mass with respect to metallicity after the first 500 Myr and clearly shows that diffusion pushes stellar metallicity towards the average, where the yield is slightly lower. Even though the total metal mass is slightly different between the runs, the differences are only small and should not affect a comparison between the runs too much.

\begin{figure}
\includegraphics[width=0.45\textwidth]{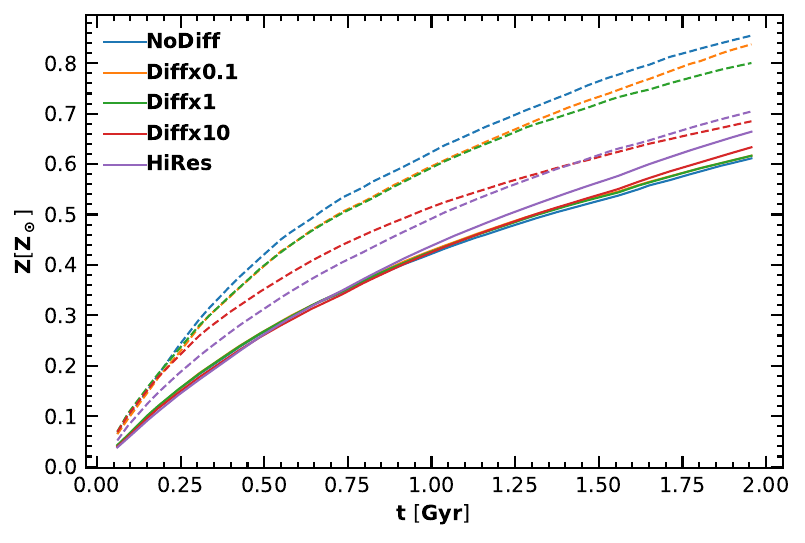}
\caption{The global (i.e. mass-weighted average) stellar metallicity (dashed) and gas metallicity (solid) as a function of time for the different models. With stronger diffusion or higher resolution, stellar metallicity is lowered and gas metallicity increased.}\label{fig:Zstar-gas}
\end{figure}

While the total mass of metals is similar in all runs, diffusion can affect how it is distributed among stars and gas. In Figure~\ref{fig:Zstar-gas}, the time evolution of the galactic stellar and gas metallicity is shown. The stellar metallicity (dashed) is generally higher than the gas metallicity (solid), but with stronger diffusion or higher resolution stellar metallicity is lowered while gas metallicity increases. If all components were perfectly mixed there would be no difference between them and so this trend is easily understood for diffusion which enhances mixing. At higher resolution, the difference between gas and stellar metallicity is even smaller. This can be attributed to the smoother distribution of sources which leads to less extreme metallicities and therefore a narrower metallicity distribution.

\subsection{Spatial Distribution of Metals and Dust}\label{sec:SpatialDistribution}

\subsubsection{Projected Distributions}

\begin{figure*}
\includegraphics[width=0.95\textwidth]{figures/spatial maps/Metallicity_005.pdf}
\caption{Density-weighted projection of metallicity of the simulated galaxies at $t = 250\,\text{Myr}$ for different runs. The top and bottom panels show the face-on view and edge-on view of the simulated galaxy, respectively. The $x$ and $y$ coordinates run from $-20$ to $20\,\text{kpc}$ and the $z$ coordinate runs from $-8$ to $8\,\text{kpc}\,$.}\label{fig:Zmap250}
\includegraphics[width=0.95\textwidth]{figures/spatial maps/Large_005.pdf}
\caption{Same as Fig.~\ref{fig:Zmap250} but for large grain abundance.}\label{fig:DLmap250}
\includegraphics[width=0.95\textwidth]{figures/spatial maps/DsDl_005.pdf}
\caption{Same as Fig.~\ref{fig:Zmap250} but for small-to-large grain ratio.}\label{fig:DsDlmap250}
\end{figure*}

\begin{figure*}
\includegraphics[width=0.95\textwidth]{figures/spatial maps/Metallicity_040.pdf}
\caption{Same as Fig.~\ref{fig:Zmap250} but at $t = 2\,\text{Gyr}$}\label{fig:Zmap2Gyr}
\includegraphics[width=0.95\textwidth]{figures/spatial maps/Large_040.pdf}
\caption{Same as Fig.~\ref{fig:DLmap250} but at $t = 2\,\text{Gyr}$}\label{fig:DLmap2Gyr}
\includegraphics[width=0.95\textwidth]{figures/spatial maps/DsDl_040.pdf}
\caption{Same as Fig.~\ref{fig:DsDlmap250} but at $t = 2\,\text{Gyr}$}\label{fig:DsDlmap2Gyr}
\end{figure*}

\begin{figure}
\includegraphics[width=0.45\textwidth]{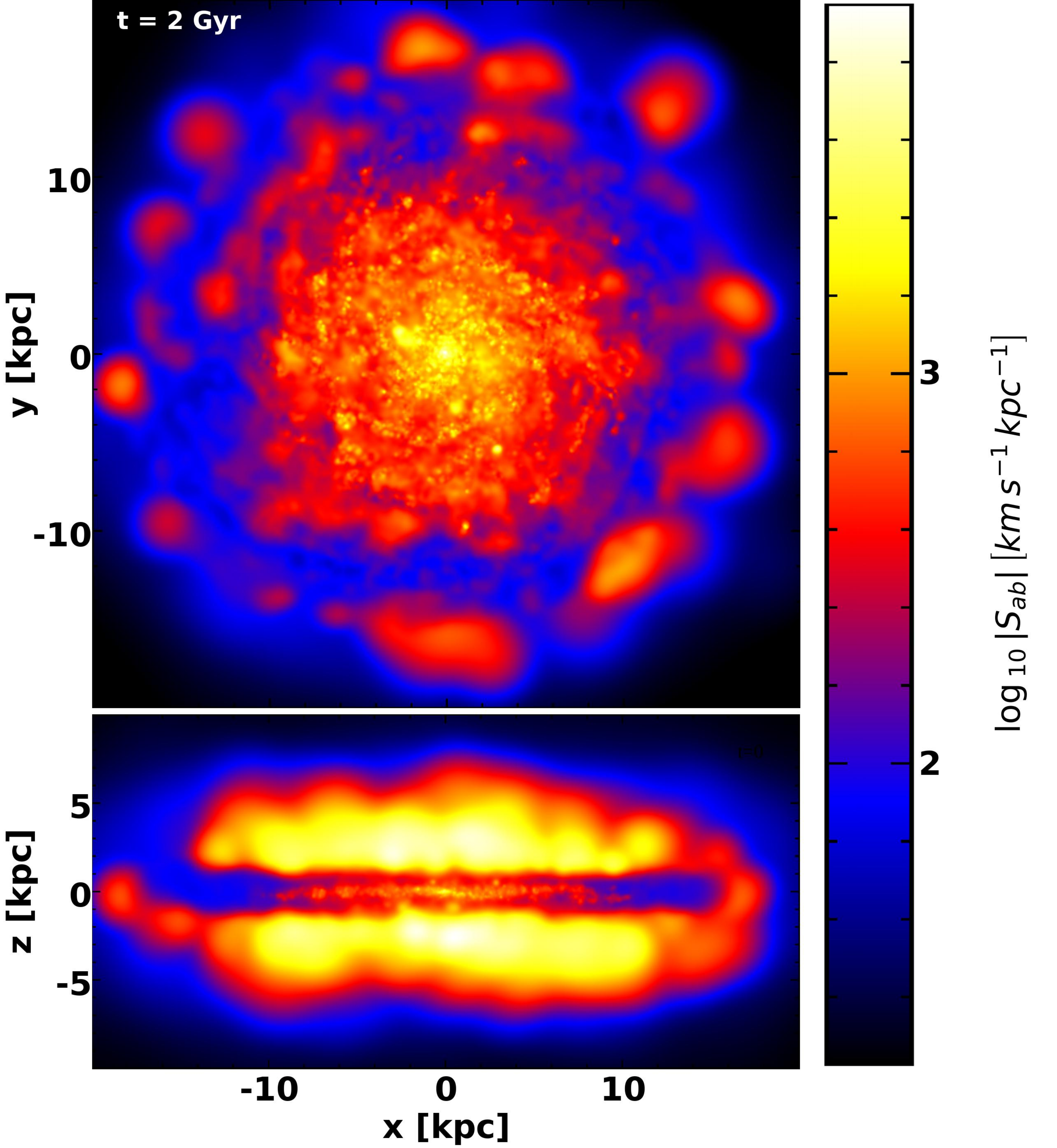}
\caption{Density-weighted projection of the shear $\left\|S_{ab}\right\|$ at $t = 2\,\text{Gyr}$ in the run with $C_{\text{d}} = 0.02$. The top and bottom panels show the face-on view and edge-on view of the simulated galaxy, respectively.}\label{fig:ShearMap}
\end{figure}

Diffusion impacts the spatial distribution of metals and dust. To highlight this, we show spatial maps of the metallicity, the large grain abundance, i.e. the mass fraction of large grains and the small-to-large grain mass ratio in Figures~\ref{fig:Zmap250} -- \ref{fig:DsDlmap2Gyr} at early ($t = 250\, \text{Myr}$) times, where metallicities are still low and we thus do not expect significant dust growth and at late times ($t = 2 \, \text{Gyr}$) and high metallicities, where the GSD has settled in its final state and is not expected to evolve much further. All projection plots have been made with the SPH visualization code SPLASH \citep{2007PASA...24..159P}.

The distribution of metallicity at early and late times is shown in Figures~\ref{fig:Zmap250} and \ref{fig:Zmap2Gyr}, respectively. As expected, the spatial distribution of metals in the run without diffusion is more grainy than in the other runs and becomes increasingly smooth as the diffusion strength is increased. At early times, the metals are extending out to similar radii, but in the run with strong diffusion they reach slightly higher altitude above and below the disk. In the run without diffusion, the metallicity hardly traces the spiral structure of the galaxy, while in all runs with diffusion it is clearly visible in the face-on view, both at early and late times, likely due to reduction in particle-particle noise.

At late times, the distribution of metals shows significantly different spatial extent among the runs. 
In the \NoDiff\ run, there are hardly any metals present beyond $r \sim 15 \, \text{kpc}$ and $\left|z\right| \sim 8 \, \text{kpc}$, while they tend to extend out further from the galactic disk with increasing diffusion strength. Furthermore, in the runs with diffusion, there is a toroidal region with low metallicity located in the galactic plane just at the edge between the disk and the halo around $r \gtrsim 15\, \text{kpc}$ which is not present in the \NoDiff\ run. We have used the particle selection criterion described in Section~\ref{sec:Selection} to check whether or not this feature is arising from the superposition of the disk with the halo, but it is present in both radial profiles and projections of the disk component, while being absent in the halo component, making it unlikely to be an artifact from the superposition of the two components. Indeed, one would expect less superposition effects with diffusion due to smoothing at the boundary. We discuss this point further in Appendix C. A likely explanation for this metal-poor torus is that the diffusion is sourced by turbulence. Within the disk, turbulence arises due to gravitational torques and stellar feedback from young stars, while in the halo it originates from random motion. In the outer parts of the disk where stellar feedback is weak
there is hardly any turbulence which could drive diffusion. This point is further illustrated in Figure\ref{fig:ShearMap}, which shows the spatial distribution of the shear $\left\|S_{ab}\right\|$ in the intermediate diffusion run at $t = 2\,\text{Gyr}\,$. The shear is strongest in the thin disk and the halo gas just above and below the disk, but is weak in the thick disk. The qualitative features of this spatial map do not differ much between different snapshots and different runs.
Thus there should be a radius within the disk beyond which diffusion is inefficient. Within the halo the fluid can easily mix and therefore will eventually populate the parts of the halo that lie in the plane of the galactic disk, leading to the formation of a metal-poor torus. Another notable detail is that, in the lower resolution runs, there are more metals in the polar regions above and below the disk than in the \HiRes\ run. This is likely due to less efficient feedback in the \HiRes\ run, owing to a lower total energy output per feedback event, which is quickly dissipated. This can lead to weaker outflows in the central region, where star formation is strongest and therefore feedback effects would become most apparent.

The spatial distribution of large dust grains is shown in Figures~\ref{fig:DLmap250} and \ref{fig:DLmap2Gyr} at early and late times, respectively. Given that a large fraction of the dust mass is locked in large grains, they are an excellent tracer of the total dust abundance and supplemented with the small-to-large grain ratio, all information about dust can be retained. Similarly to the distribution of metallicity, the distribution of large grains is increasingly smoother with stronger diffusion and traces the spiral structure of the galaxy, once diffusion is included. The distribution of large grains closely follows the overall distribution of metals within the disk, but rapidly drops with the onset of the halo. This is contrary to the na\"ive  expectation, that the halo might in fact be a rather hospitable environment for large grains. Shattering due to grain-grain collisions, which most efficiently reduces the large grain abundance, is rather inefficient in the halo, due to the extremely low densities. Furthermore, since large grains have a low surface-area-to-volume ratio, thermal sputtering, which efficiently destroys small grains in the halo, can only reduce the number of large grains relatively slowly. 
In spite of this, the fact that there is only a low abundance of large grains in the halo, owes to their production in the dense environment of the galactic disk. In order to reach the very diffuse parts of the halo, large grains need to travel through gas with intermediate density, where they are shattered efficiently. As a result, hardly any large grains make it into the halo, while the shattered fragments are destroyed through thermal sputtering. Interestingly, hot and diffuse outflows, i.e. due to strong SN or AGN feedback, might constitute a way for large grains to escape from the disk and populate the halo.

In order to see where dust processing is most efficient, the spatial distribution of the small-to-large grain ratio \DsDl is shown in Figures~\ref{fig:DsDlmap250} and \ref{fig:DsDlmap2Gyr} at early and late times, respectively. At early times, enrichment with small grains is only seen in the spiral arms of the disk and it extends out to larger radii in the \NoDiff\ and \WeakDiff\ runs. With stronger diffusion, small grains are initially restricted to the very center of the disk and cannot reach far from the relatively metal-rich center. These differences are likely explained by the fact that metal mixing initially leads to a \emph{dilution} phase, characterised by a mixing timescale that is shorter than the growth timescale. During this phase local variations in the metal and dust field are smoothed out, resulting in lower peak metallicities. Once the metallicity in the well mixed gas is large enough, growth starts, leading to a sharp rise in the small grain abundance. Without diffusion, whichever cells get enriched with metals first can grow first, leading to a headstart in growth compared to the rest of the cells, explaining the overall higher small grain enrichment at early times.

At late times, in all runs the full galactic disk is enriched with small grains, and \DsDl increases from the center towards the edge of the disk and then falls off towards the halo. The exact radius at which the maximum is attained seems to be rather independent of the strength of diffusion as long as some degree of diffusion is present. Within the halo, in the vicinity of the disk \DsDl is higher in the runs with stronger diffusion. Furthermore, in the runs with diffusion \DsDl becomes large within a thin layer above and below the thin disk, whereas it tends to fall off with increasing height in the \NoDiff\ run. In the thin disk, log(\DsDl)$\gtrsim 0.5$ corresponding to the thin yellow strip in the midplane. Just above and below the midplane, in the thick disk in all runs with diffusion log(\DsDl) is closer to 0 corresponding to the red envelope that extends up to $(1 - 2)\, \text{kpc}$, before it falls off in even higher layers. This is due to a steady supply of large grains from the thin disk into the surrounding layers, which can be efficiently shattered, leading to the slightly increased \DsDl.

\subsubsection{Radial Profiles}\label{sec:profiles}

\begin{figure}
\includegraphics[width=0.45\textwidth]{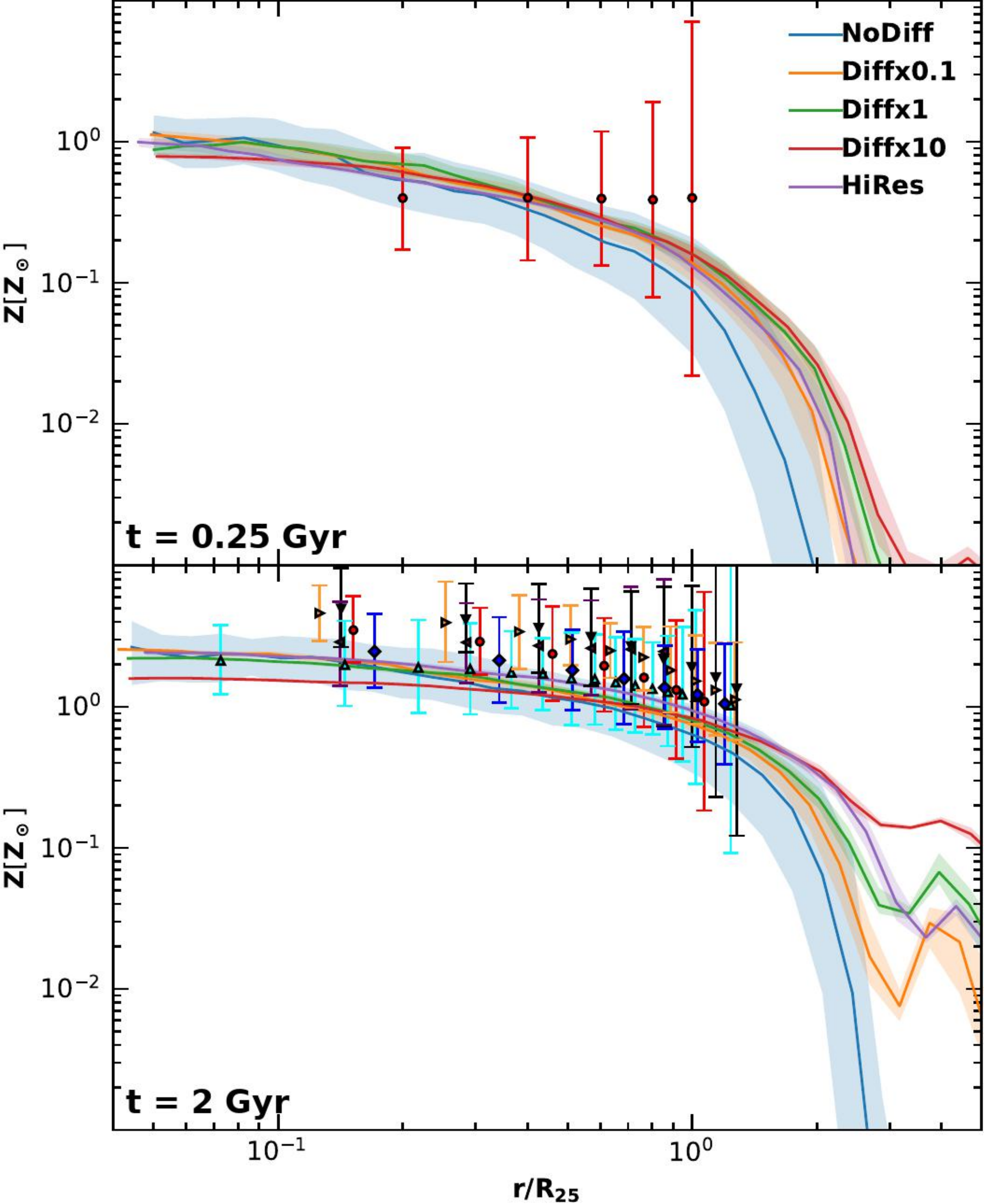}
\caption{The gas metallicity as a function of cylindrical radius $r$ for the different simulations at early (top) and late (bottom) times. We normalised $r$ by $R_{25}$(see text). Solid lines depict the median relation, while the shaded area shows the range of values between the 25th and 75th percentile. We compare our results with observational data compiled by \citet{2012MNRAS.423...38M}. In the top panel the filled circles with errorbars correspond the data in Holmberg II. In the bottom panel the circle, $\triangle$, $\triangleleft$, $\triangleright$, $\triangledown$ and diamond symbols correspond to NGC628, NGC2403, NGC4736, NGC5055, NGC5194, NGC7793, respectively.}\label{fig:RadialZ}
\end{figure}
\begin{figure}

\includegraphics[width=0.45\textwidth]{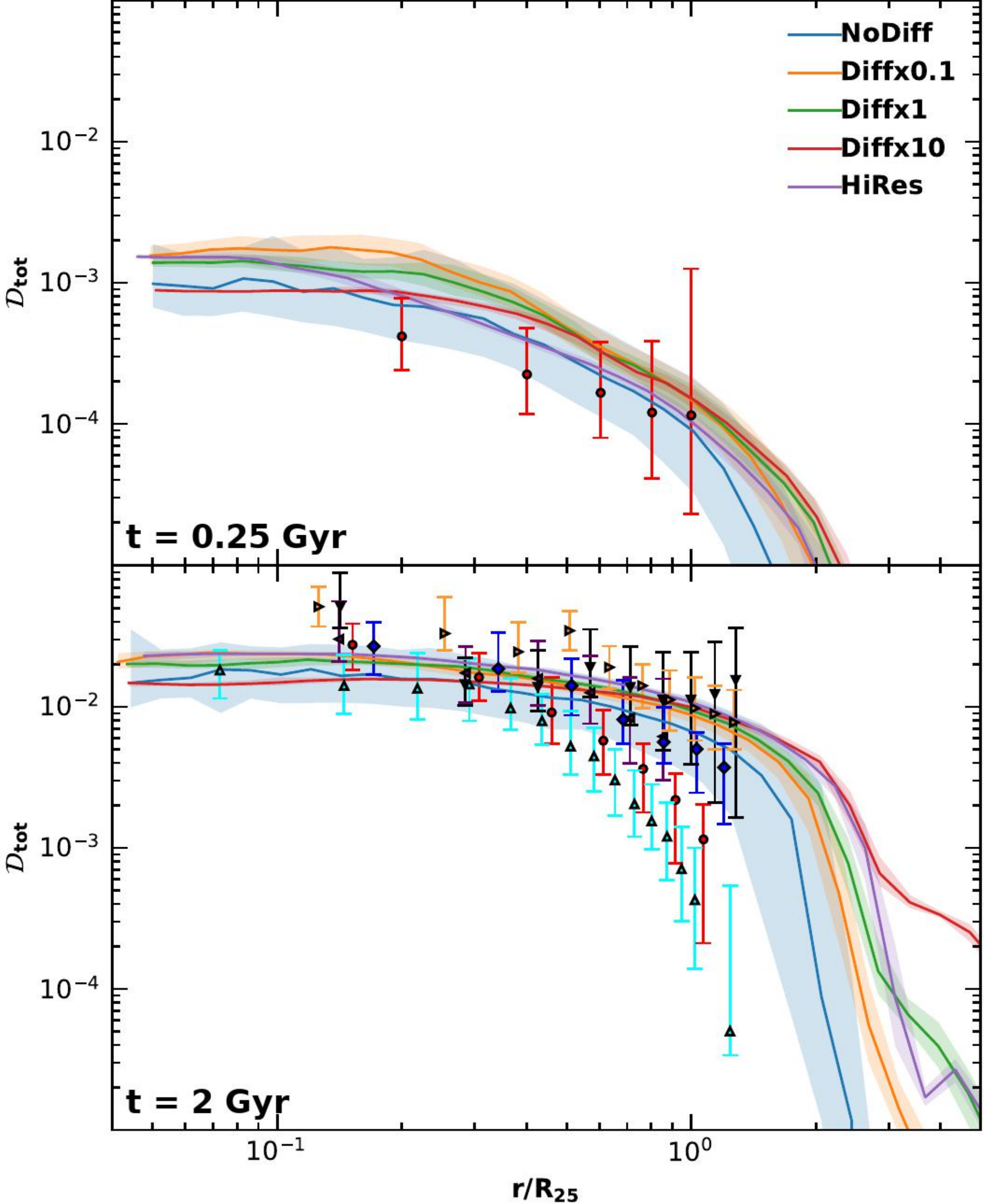}
\caption{Same as Fig. \ref{fig:RadialZ} but presenting the radial profile of the dust-to-gas ratio.}\label{fig:RadialDust}
\end{figure}

In Figures~\ref{fig:Zmap250} to \ref{fig:DsDlmap2Gyr}, we have seen that metal and dust components of the galaxy follow qualitatively similar radial trends. We have also shown that diffusion leads to the formation of a metal-poor torus in galactic plane. In order to quantitatively assess the impact of diffusion and compare the models to the observations, in this section we present radial profiles of certain quantities like dust abundance and metallicity. To this end, we assign particles to logarithmically spaced radial bins. For each bin, we compute the median, 25th and 75th percentile of the quantities of interest. We normalize the radius to $R_{25}$ (the radius beyond which the surface brightness falls below 25 mag arcsec$^{-2}$), in order cancel the galaxy size effect, enabling us to compare our results to observations of several spiral galaxies at once.
We evaluate $R_{25}$ for the simulated galaxies, by using the relation $R_{25} \simeq 4 R_{\text{d}}$ \citep{1998ggs..book.....E}. Here $R_{\text{d}}$ refers to the scale length of the radial column density profile of young stars, obtained by fitting them with an exponential function. In our simulated galaxies $R_{25}$ grows from $R_{25} \sim 4\,\text{kpc}\,$ at early times to $R_{25} \sim 4.5\,\text{kpc}\,$ at late times with mild differences between each run.
 
In order to allow for a better comparison with previous theoretical work, we overplot some of the profiles with the same observational dataset as the one adopted by \citetalias{2017MNRAS.466..105A} and \citetalias{2020MNRAS.491.3844A} which has been compiled by \citet{2012MNRAS.423...38M}. 
The data consists of spatially resolved data of the dust-to-gas ratio and the dust-to-metal ratio in a sample of nearby galaxies chosen from the \emph{Spitzer} Infrared Nearby Galaxies Survey sample \citep{2003PASP..115..928K}. \citetalias{2017MNRAS.466..105A} categorised the sample by classifying them according to their specific star formation rates, allowing to map them to different simulation snapshots. We apply their classification in order to perform a similar mapping. We compare the dwarf irregular galaxy Holmberg II to the snapshots at $t = 250\,\text{Myr}\,$ and the galaxies NGC628, NGC2403, NGC4736, NGC5055, NGC5194, NGC7793 to the snapshots at $t = 2\,\text{Gyr}\,$. We note, that \citet{2012MNRAS.423...38M} used a metallicity calibration that might overestimate metallicities, in order to avoid dust-to-metal ratios exceeding unity. This might bias the observed values of \DtoZ to lower values.

\begin{figure}
\includegraphics[width=0.45\textwidth]{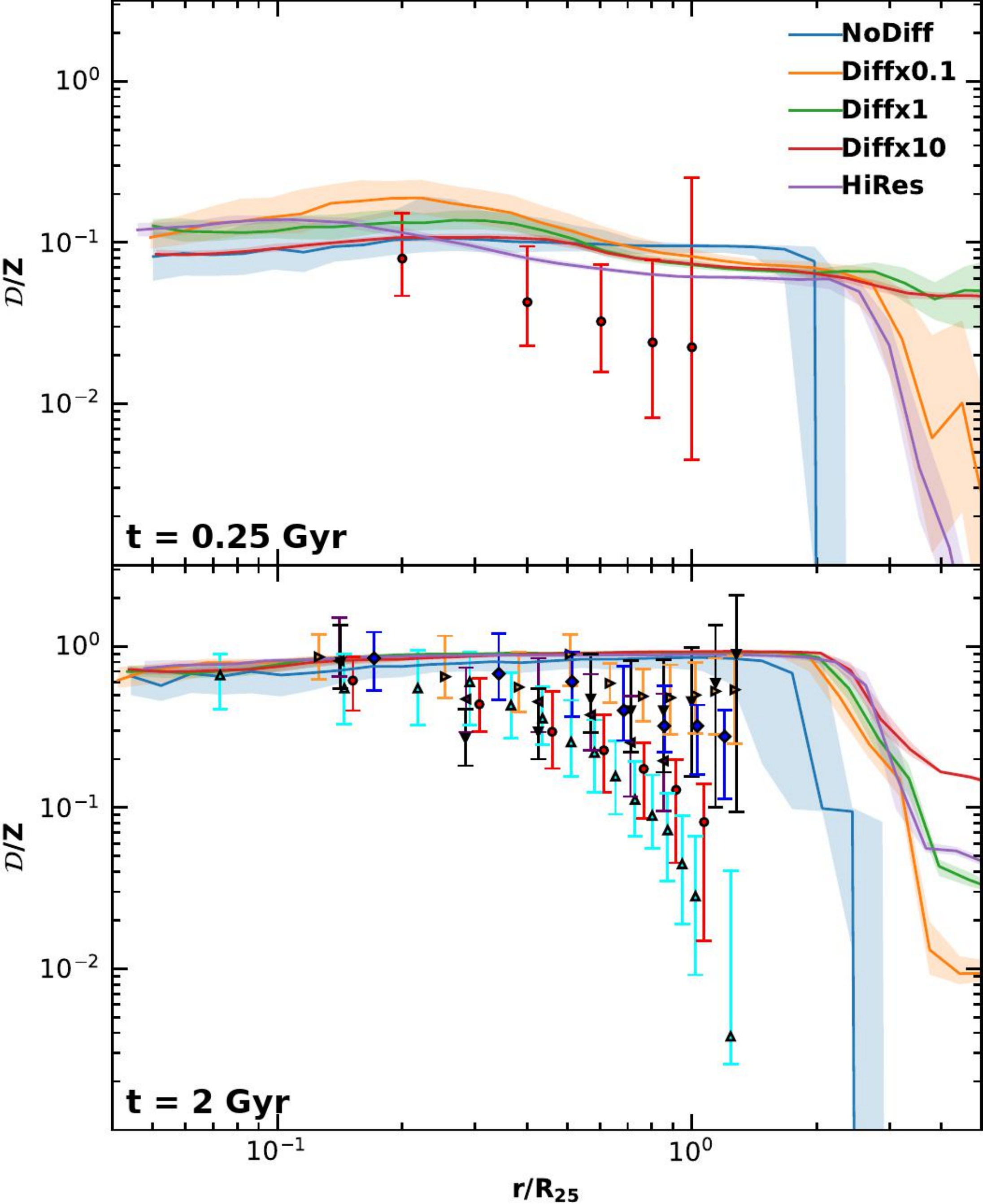}
\caption{Same as Fig. \ref{fig:RadialZ} but presenting the radial profile of the dust-to-metal ratio.}\label{fig:RadialDZ}
\end{figure}

In Figure~\ref{fig:RadialZ}, the radial metallicity profile at early ($t = 250\,\text{Myr}$) and late ($t = 2\,\text{Gyr}$) times is shown for the different models. Observational data points for $Z$ are derived from the dataset from \citet{2012MNRAS.423...38M}. In all models the metallicity is peaked at the center and gradually falls off towards $r \sim R_{25}$, beyond which it rapidly drops. Without diffusion, the dispersion of metallicity values is generally larger than in the runs with diffusion. This is generally true, as diffusion leads to the  local averaging of diffused quantities pulling their values closer to the local mean value. At early times and small radii, all models agree quite well with each other and differences only become visible at large radii, where the radius beyond which the metallicity starts to drop rapidly increases with diffusion strength. In contrast, the observational data (Holmberg II) exhibits no significant metallicity gradient. This difference might be due to the different morphology of Holmberg II, which is classified as a dwarf irregular galaxy. Nonetheless, overall the metallicity levels agree with our simulations. At late times, the drop at the edge of the disk is shallower with stronger diffusion, as diffusion acts to populate the halo with metals. In the runs with diffusion, the metallicity reaches a local minimum at $r \gtrsim 3 R_{25}$ before increasing towards a local maximum at $r \lesssim 4 R_{25}$ beyond which it falls off rapidly. The minimum is more pronounced with weaker diffusion, but requires some degree of diffusion to be present in the first place. In the \HiRes\ run, the minimum is less pronounced than in the \WeakDiff\ run, but more pronounced than in the \Diff\ run, indicating that the effective diffusion strength is somewhere in between the two. This region around the local minimum corresponds to the metal-poor torus mentioned in the discussion of Figure~\ref{fig:Zmap2Gyr} above. Very strong diffusion ($C_d = 0.2$) acts to flatten the metallicity profile and reduces the slope of the drop in metallicity outside the disk. At lower values of the diffusion strength, the slopes in the central region and the outer parts of the halo are hardly changed compared to the case of no diffusion, but the profile is slightly shallower at intermediate radii $r \sim R_{25}$. The observational data tend to lie above the simulation data. Since metallicity tends to increase with time, this might indicate that a comparison with a later snapshot might lead to better agreement. Nonetheless, apart from the normalization, the overall radial trend is well captured by all models. The runs with diffusion tend to be in better agreement with the data at large radii, where the profiles tend to be flatter with diffusion. 

To our knowledge, the metal-poor torus seen in the runs with diffusion has not been observed in spatially resolved observations of nearby spirals. Here we want to explore why this might be. First it should be noted that spatially resolved observations of metals and dust require bright sources, i.e. it is significantly easier to make reliable spatially resolved observations of the dense inner parts of a galaxy, while it is exponentially more difficult and significantly less conclusive if one tries to make similar observations of the outer part of the disk or the halo. This is why most spatially resolved observational data do not extend much beyond $R_{25}$. Given that the metal-poor torus is located at around $r\sim 3-4 R_{25}\,$, it is thus likely that current observational methods are simply not able to reasonably resolve such a structure and even in the case that observations at this radius existed, the scatter in the observed metallicity values might be larger than the depth of the dip in metallicity, i.e. such a torus would be unobservable. 
Furthermore, it is possible that the torus is just a numerical relic arising from the perfectly spherical structure of the halo and the absence of environmental effects like mergers, filaments or major in- or outflows. It is thus unclear, whether such a torus would even form in the first place, in the more realistic cosmological environment, which includes all of these disruptive features.

Figure~\ref{fig:RadialDust} depicts the radial profile of the dust-to-gas ratio \Dtot at early and late times. Observational data are taken from \citet{2012MNRAS.423...38M}. The dust profile exhibits similar features as the metallicity profile, but at early times is flatter in the center and at late times exhibits a less pronounced feature at the location of the metal-poor torus in the runs with diffusion. Central dust abundances are slightly higher with diffusion than without, except in the \StrongDiff\ run where a similar abundance is reached. In all runs, the central slope flattens over time as the disk becomes more enriched with dust, while the slope at the edge of the disk steepens. At early times all models agree well with the data in Holmberg II; however overall levels of \Dtot\ tend to be slightly lower in Holmberg II. At late times, the observational data have a large dispersion. All simulations tend to agree reasonably well with most profiles, though the runs with diffusion, which tend to have a flatter and more extended profile are in better agreement with the data in NGC5055, NGC5194 and NGC7793, while NGC628 and NGC2403 where the profiles fall off more rapidly tend to be more consistent with the run without diffusion.
The results without diffusion are in reasonable agreement with \citetalias{2020MNRAS.491.3844A}'s previous simulation without diffusion.

\begin{figure}
\includegraphics[width=0.45\textwidth]{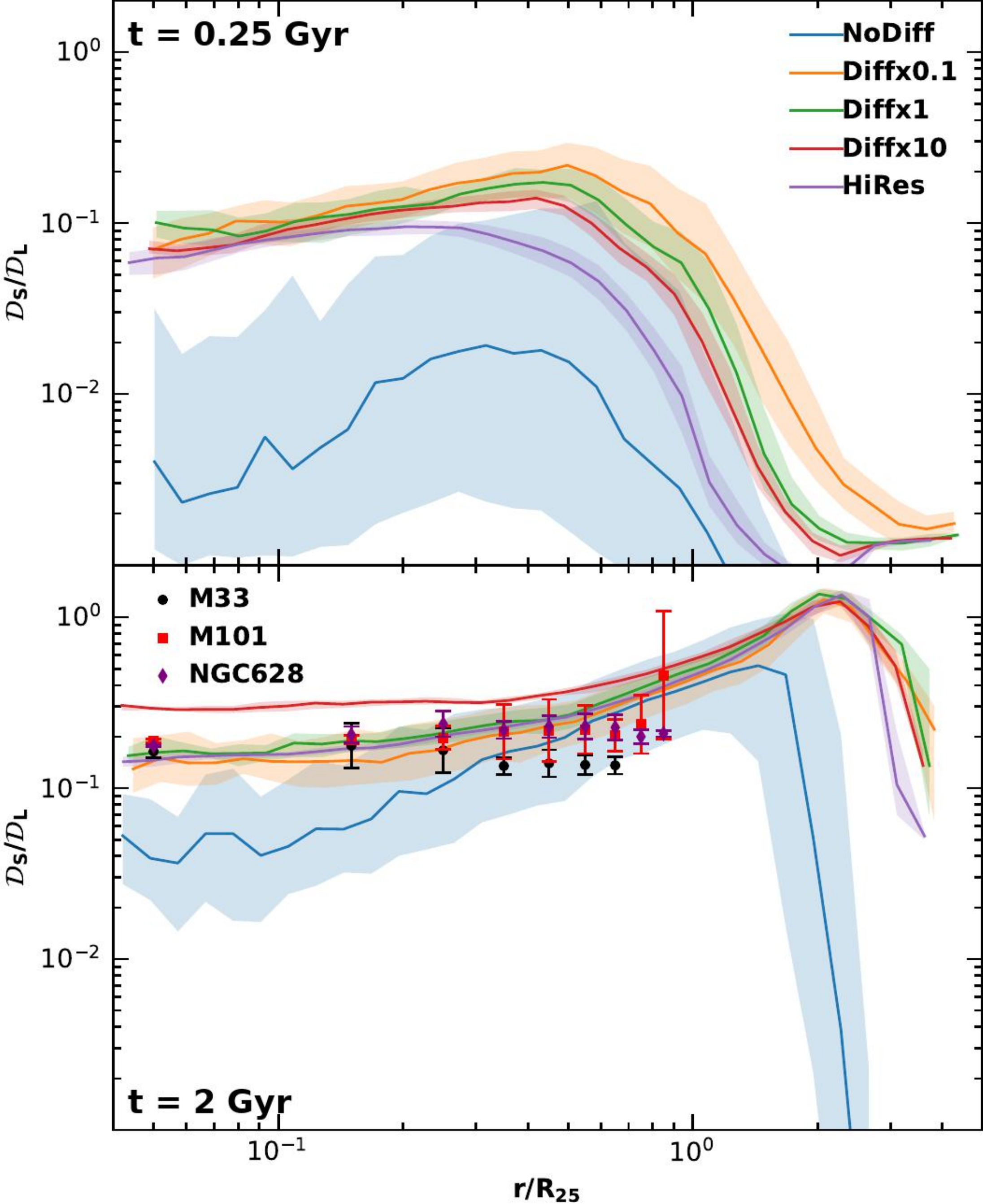}
\caption{Same as Fig. \ref{fig:RadialZ} but presenting the radial profile of small-to-large grain ratio. The data points are taken from \citet{2020A&A...636A..18R}. The galaxies from this sample are late stage spirals, which is why we are only comparing them to the snapshot at late times.}\label{fig:RadialDsDl}
\end{figure}

Figure~\ref{fig:RadialDZ} shows the radial profile of the dust-to-metal ratio \DtoZ at early and late times. Observational data are taken from \citet{2012MNRAS.423...38M}. At early times, there are major differences between the models. In the \NoDiff\ run, \DtoZ is roughly constant at \DtoZ$\sim 0.1$ within the disk, but rapidly drops beyond $r \gtrsim 2 R_{25}$.
In the other runs, it is slightly larger in the center but drops to \DtoZ$ \sim 7 \times 10^{-2}\,$ beyond $r \gtrsim 0.3 - 0.5 R_{25}$. The central increase in \DtoZ is larger with weaker diffusion. With strong diffusion, there is no increase in the central value of \DtoZ, but at intermediate radii it is still lower than without diffusion. In the runs with diffusion, the drop in \DtoZ beyond $r \sim 2 R_{25}$ is shallower and happens at larger radii than in the \NoDiff\ run. The profile in the \HiRes\ run is very different from those in the \Diff\ and \StrongDiff\ runs, indicating that \DtoZ might be sensitive to the details of the feedback prescription. In all runs, the values of \DtoZ tend to be higher than the observed values in Holmberg II. \DtoZ tends to decrease with increasing radius, a trend which is captured in the runs with diffusion, slightly improving the agreement compared to the run without diffusion, where \DtoZ is constant throughout the disk. 

The radial trend of the dust-to-metal ratio at early times may be explained by the fact that the galaxy is still actively star forming, which leads to stronger feedback and efficient outward advection of dust- and metal-enriched particles. In the \NoDiff\ run, this leads to the build-up of a constant \DtoZ profile consistent with the yield relation. In the runs with diffusion, the large grains diffuse into the outer parts of the disk where they can be shattered, leading to the formation of small grains. At this point, the gas in the center is still hosting mostly large grains and thus diffusion transports some of the small shattered fragments to the central region.
The small grains accrete gas-phase metals, so that the dust-to-gas ratio in the center slightly increases.
However, the central star-forming part of the disk is a very hostile environment for small grains at early times, since sputtering due to SN explosions efficiently consumes their abundances. Thus the overall destruction rate of small grains in the central region increases with increasing diffusion rate, resulting in lower \DtoZ in the central region.
Since the only sources of dust grains in the outer parts of the disk are advection and diffusion of large grains from the center, this leads to lower dust-to-metal ratios in this region.

At late times, the feedback is more quiescent and differences between the simulations arise mostly due to differences in grain growth. All runs exhibit a similar \DtoZ profile within $r < 2 R_{25}$, which is rather flat at around \DtoZ $\sim 0.8 - 0.9$, with a slight decrease towards small $r$. The profile in the \NoDiff\ run reaches slightly lower values than the other runs and drops steeply just beyond $r \sim 2 R_{25}$. In the runs with diffusion the steep drop only happens at slightly larger radii. With strong diffusion, higher values of \DtoZ exceeding the yield relation \DtoZ$ = 0.1$ are attained even in the halo, indicating the presence of large grains, as small grains would be lost to sputtering almost immediately. At late times, the values of \DtoZ from the simulations tend to be slightly higher than the observational data at large radii. This is expected, since we are neglecting the effect of dust composition, which leads to an overestimate of metals which can be readily condensed into dust grains. None of the models can explain the steep decline in \DtoZ in the systems NGC628 and NGC 2403. \citetalias{2020MNRAS.491.3844A} report the same issue of too flat dust-to-metal ratio profiles in their simulation. They discuss that this may be related to too efficient accretion, which may be related to the missing distinction between different grain species. 

In Figure~\ref{fig:RadialDsDl} the radial profile of the small-to-large grain ratio is shown at early and late times. Observational data points are taken from \citet{2020A&A...636A..18R}. They fitted near to far-IR maps of
the three spiral galaxies M33, M101 and NGC628 on a pixel-by-pixel basis and derived dust maps within the disks of their galaxy sample. They fit the data with the classical dust model by \citet{1990A&A...237..215D}, assuming three types of grains: polycyclic aromatic hydrocarbons (PAHs) and very small graphite grains for the small grains and big silicates ($a > 0.15 \,\micron$) for the large grains. Such a comparison is bound to exhibit certain differences due to numerous uncontrolled systematic effects related to the exact evolutionary stage and assembly histories of the galaxies; however, it is still worthwhile and might lead to some insight.

In all runs, the early time \DsDl profile exhibts a gentle increase from the center towards intermediate radii, where a local maximum is attained. Beyond the maximum, \DsDl drops rather mildly and, in the runs with diffusion, reaches a constant value of $\left(\mathcal{D}_{S}/\mathcal{D}_{L}\right)_h \sim 2 \times 10^{-3}\,$. The shape of the profile is significantly altered by the inclusion of diffusion. In the \NoDiff\ run the maximum value of \DsDl $\sim \mathcal{O}\left(10^{-2}\right)$ is attained at $r \sim 0.2 R_{25}$. If diffusion is enabled, the central value of \DsDl $\sim 0.1$ is independent of the diffusion strength, while the exact value and the position of the maximum depend on the strength of diffusion. With weaker diffusion, the maximum tends to be further out at larger radii and reach slightly larger values, even though the variation is small as $\left(\mathcal{D}_{S}/\mathcal{D}_{L}\right)_\text{max} \sim 0.1 - 0.2$. In the \Diff\ and \StrongDiff\ runs the drop from the maximum to $\left(\mathcal{D}_{S}/\mathcal{D}_{L}\right)_h$ is almost identical, with \DsDl being slightly lower in the case of stronger diffusion, while the drop is shallower in the run with weaker diffusion. The \DsDl profile in the \HiRes\ run is almost constant below $r \sim 0.3 R_{25}$ and then falls with a slope similar to the \Diff\ and \StrongDiff\ runs, even below $\left(\mathcal{D}_{S}/\mathcal{D}_{L}\right)_h$. At $r \sim 2 R_{25}$ the small-to-large grain ratio reaches a local minimum slightly below \DsDl$\sim 10^{-3}$ and increases from there towards $\left(\mathcal{D}_{S}/\mathcal{D}_{L}\right)_h$.

At late times, the shape of the profile has changed significantly, and in the central region differences between the runs with and without diffusion have become smaller. In the \NoDiff\ run, the central value of \DsDl has increased compared to early times to \DsDl$\sim 6 \times 10^{-2}\,$ and the profile mildly increases up to $r \sim 2 R_{25}$ where it reaches its maximum at $\left(\mathcal{D}_{S}/\mathcal{D}_{L}\right)_\text{max} \sim 0.3$. Beyond the maximum, the small-to-large grain ratio drops steeply. In the runs with diffusion the central value of \DsDl$\sim 0.1 - 0.2$ is rather independent of the diffusion strength, but with strong diffusion the central value is $2 - 3$ times larger. In the runs with diffusion, \DsDl is almost constant and only mildly increases up to $r \lesssim R_{25}$ where the slope slightly steepens. At $r \sim 2.5 R_{25}$ a maximum value of $\left(\mathcal{D}_{S}/\mathcal{D}_{L}\right)_\text{max} \sim 1$ is reached. At larger radii the profile drops towards an almost constant value of \DsDl that is higher in the runs with stronger diffusion. At small radii, the \WeakDiff, \Diff\ and \HiRes\ runs agree well with the observations. The \StrongDiff\ and \NoDiff\ runs are exhibiting too high and too low values of \DsDl, respectively. At intermediate radii the discrepancy becomes less prominent.
However none of the models reproduce the mild drop in \DsDl towards intermediate radii. Possible reasons for this discrepancy are manifold. Environmental effects like mergers or cold gas inflows which are not present in the simple case of an isolated galaxy can cause radial inflows which can impact the morphology of the galaxy at all radii and change how metals and dust are distributed within the galaxy \citep{2014MNRAS.438.1870D}. Furthermore, some of the the assumptions made in the dust model used to fit the observations like optically thin emission \citep{1998ApJ...509..103S} may bias the results.

\subsection{Evolution of the Grain Size Distribution}

The GSD is determined by two main components: its normalization, i.e. the dust-to-gas ratio, and its shape, as captured by the small-to-large grain ratio. Here we will study how these two properties of the GSD evolve with time and metallicity. We also show how the full GSD in the dense and diffuse ISM changes over time. 

\subsubsection{Global Picture}

\begin{figure}
\includegraphics[width=0.45\textwidth]{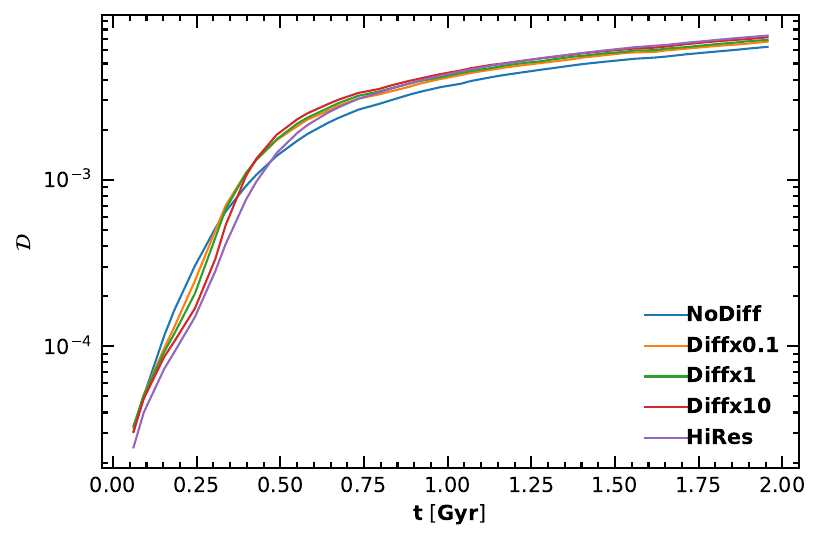}
\caption{The galactic dust-to-gas ratio as a function of time for the different models.}\label{fig:DtotHistory}
\end{figure}

\begin{figure}
\includegraphics[width=0.45\textwidth]{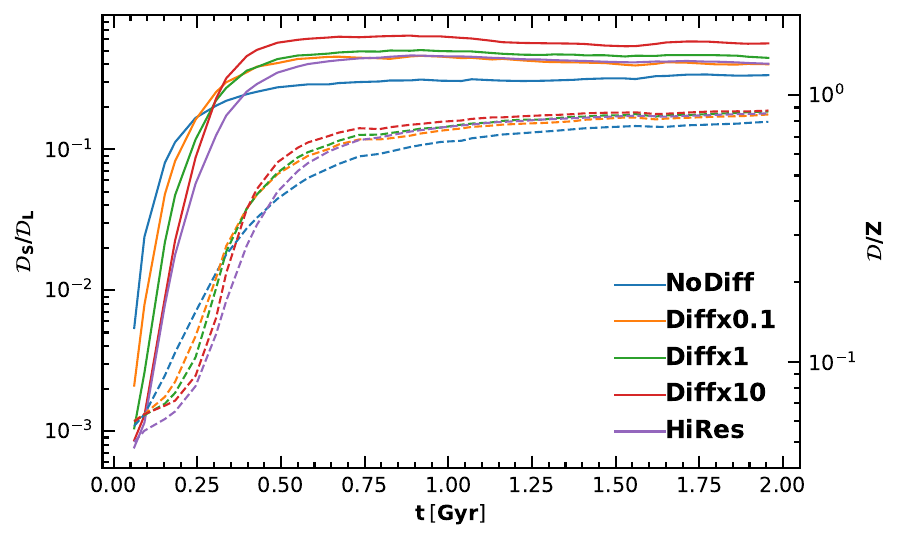}
\caption{The galactic small-to-large grain ratio (solid) and the dust-to-metal ratio (dashed) as a function of time for the different models.}\label{fig:DsDlHistory}
\end{figure}

In order to compare the efficiency of overall dust growth in the models, we study the evolution history of the dust-to-gas ratio, shown in Figure~\ref{fig:DtotHistory}. In all models, the dust abundance first undergoes a phase of exponential growth, before it slows down at around $t = 500\,\text{Myr}\,$ after which the dust abundance only grows linearly, limited by the very slow injection of new metals. The initial growth is slower, but longer in the runs with stronger diffusion. The final value of the dust-to-gas ratio is higher by about $16\,\%\,$ with stronger diffusion.

Comparing this trend to the evolution of the small-to-large grain ratio might give some insight into what drives the dust growth. The global small-to-large grain ratio is defined as the ratio of the total mass of small and large grains within the disk. The time evolution of \DsDl and \DtoZ is shown in Figure~\ref{fig:DsDlHistory} for the different models. In all runs, \DsDl first undergoes a phase of exponential growth, before it saturates, similar to the time evolution of \Dtot. The time evolution of \DtoZ reflects the trends seen in the time evolution of \DsDl, but is slightly offset, indicating that grain growth occurs only after grain processing shifted the GSD towards smaller grain sizes.
There are a number of notable differences in the evolution histories of the different models. In the runs with weaker or no diffusion, \DsDl initially grows at a higher rate, but for a shorter time, resulting in overall lower values of \DsDl and \DtoZ at saturation. The differences in the growth phase due to diffusion can be explained by noting that diffusion initially leads to a dilution phase, delaying grain growth. As long as the diffusion timescale is shorter than the growth timescale, newly grown small grains tend to be shared among more gas particles, leading to a dilution of grain abundances. Since the grains tend to grow faster in regions with higher abundances, this dilution initially slows down overall levels of grain growth. However, once sufficiently high abundances are reached, since there are more gas particles involved, the growth can go much further, leading to more growth in the long run. In the \HiRes\ run, \DsDl initially grows as slowly as in the \StrongDiff\ run and after a few $100 \,\text{Myr}\,$, the growth slows down even further and saturation sets in slightly delayed at a similar value as in the \WeakDiff\ run.     
This is in line with the above discussion of \DtoZ at early times. If at high resolution, there is more turbulence at early times, this would mean that diffusion can be comparable, and even stronger than in the run with strong diffusion, delaying the growth of small grains, until the feedback driven turbulence has dispersed and diffusion weakens to a level, that is comparable to the runs with intermediate or weak diffusion. In the runs with diffusion, the final value of \DsDl is higher by a factor of 1.5 to 2 compared to the \NoDiff\ run, indicating that diffusion enhances the production of small grains. 

There are three interesting time intervals at which the GSD is expected to be qualitatively different. At early times ($t \sim 250\,\text{Myr}\,$), we expect the GSD to be yield dominated, i.e. closely following the log-normal distribution from the stellar yield relation. At intermediate times ($t \sim 750\,\text{Myr}\,)$, we expect the abundance of small grains to have reached its maximum. Finally at late times ($t \sim 2 \,\text{Gyr}\,$), we do not expect the GSD to change much. In the following, we will analyse the GSD at these points in time and compare the effect of the different models at each point in time.

\subsubsection{Dust-to-gas Ratio vs. Metallicity}

\begin{figure}
\includegraphics[width=0.45\textwidth]{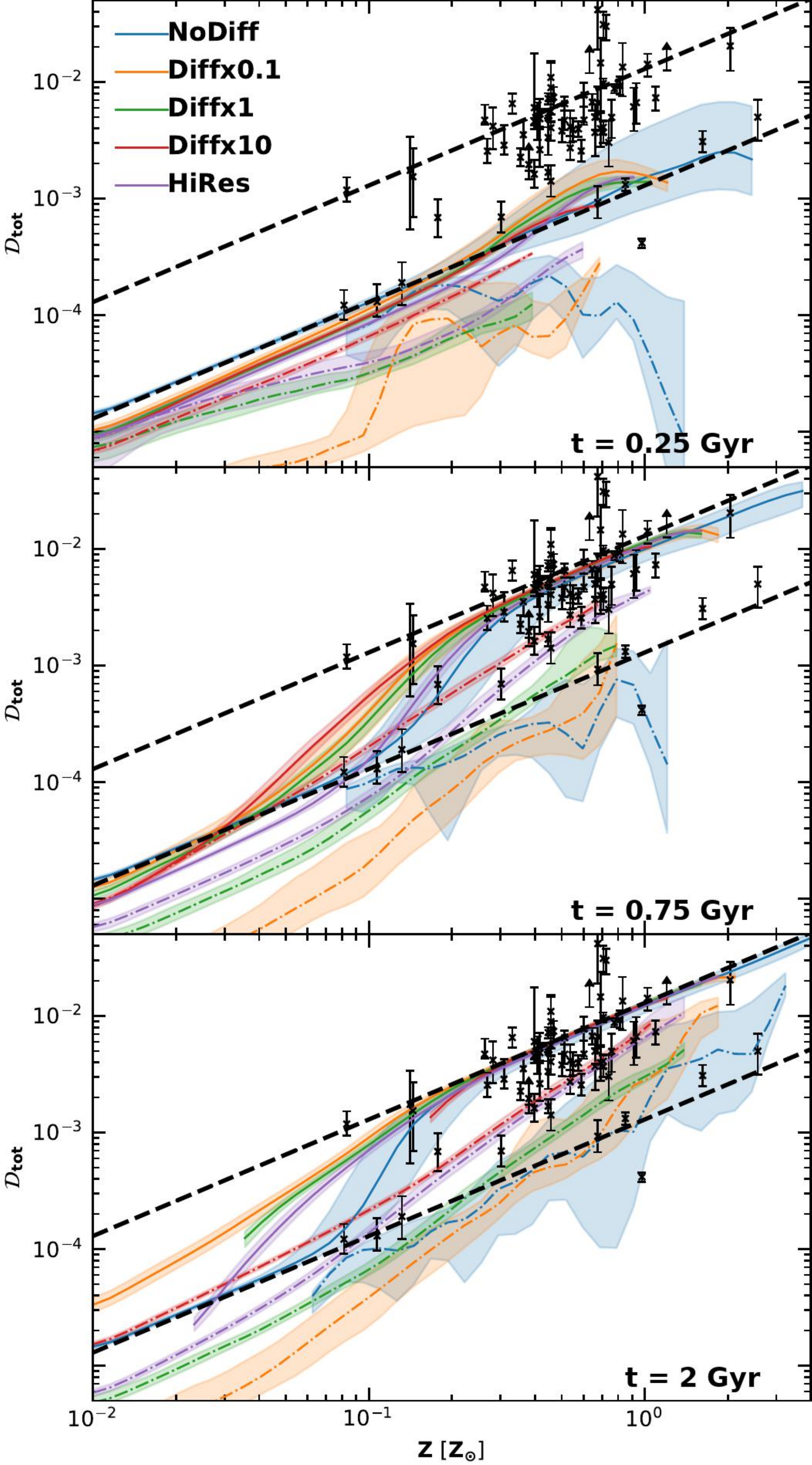}
\caption{The dust-to-gas ratio as a function of metallicity at early (top panel), intermediate (center panel) and late times (bottom panel) for the different runs. The solid lines show the median relation in the disk, while dashed lines show the relation in the halo. The shaded area corresponds to the range between the 25th and 75th percentile. The black dashed lines correspond to the yield relation \Dtot$ = 0.1 Z$ and saturation \Dtot$ = Z$, respectively. We compare our results to the observational data from \citet{2014A&A...563A..31R}.}\label{fig:DtotvsZ}
\end{figure}

We use the relation of the dust-to-gas ratio, \Dtot, with metallicity, $Z$, in order to verify that our model reasonably calculates the dust abundance. This relation is often used as a benchmark to assess the validity of chemical evolution models \citep[e.g.][]{1998ApJ...496..145L, 2020MNRAS.491.3844A}. In Figure~\ref{fig:DtotvsZ}, we show the relation between \Dtot and $Z$ in the early, intermediate, and late stage of the evolution of the galaxy for each run. We show separately the relation in the disk (solid lines) and halo (dashed lines) by grouping the particles according to the criterion in Eq.~(\ref{eq:selection}). We also plot the observed \Dtot--$Z$ relation for a sample of nearby individual galaxies compiled by \citet{2014A&A...563A..31R} from the KINGFISH survey and the sample from \citet{2011A&A...532A..56G}. The conclusions derived from the comparison are not affected even if we use newer analysis \citep[e.g.][]{2019A&A...623A...5D,2021A&A...649A..18G} for the observational data. The simulation data correspond to the median relation of gas particles within a single galaxy. Thus a direct comparison cannot be made. Instead, we just use the observational data as a first reference to decide whether or not our model produces results in line with the observed relation. All models are roughly in line with the observations, indicating that our implementation reasonably reproduces the trend of dust evolution with metallicity.

In the disk, the relation is generally characterised by \Dtot following the linear yield relation at low $Z$ and saturation at high $Z$. At intermediate metallicities non-linear growth kicks in connecting the two regimes \citep[see, e.g.][]{1998ApJ...501..643D, 1999ApJ...510L..99H, 1999ApJ...522..220H, 2003PASJ...55..901I, 2008A&A...479..453Z, 2013EP&S...65..213A, 2020MNRAS.491.3844A}. At high metallicity, the dust-to-gas ratio tends to slightly drop, which is likely related to sputtering due to SN shocks, as particles with high $Z$ tend to be located closer to stars. In the halo, typically lower dust abundances are achieved, due to thermal sputtering.

At early times, the dust has not yet experienced growth in the ISM; thus, the dust-to-gas ratio in the disk closely follows the stellar yield relation in all runs. In the runs with diffusion, \Dtot falls below the yield relation at low $Z$. This is because the gas in this regime is typically in the warm phase, where large grains can be shattered, leaving behind small fragments. Shattering increases the small grain abundances in the diffuse ISM to levels, that are larger than in the dense ISM, leading to an outflow of grains from the diffuse ISM that lowers the total dust-to-gas ratio. This is further enhanced by the erosion of small grains due to thermal sputtering in the halo and destruction in SN shocks in the dense ISM.
Diffusion also leads to higher dust-to-gas ratios in the halo. These effects are more pronounced with stronger diffusion.

At intermediate times, dust growth has started to saturate at high metallicity. In all runs, the \Dtot--$Z$ relation in the disk exhibits yield dominated low $Z$ gas and dust saturated high-$Z$ gas, with a transition region around $Z \sim 0.1 Z_{\sun}$. 
Just as at early times, the relation in the runs with diffusion lies slightly below the yield relation at low $Z$, which again is due to the excess of small grains which are easily destroyed in the halo and star forming regions and therefore keep flowing out of the diffuse ISM. Diffusion lowers the metallicity at which non-linear growth kicks in, through mixing of enriched high $Z$ gas where grain growth has already commenced and lower metallicity gas, but keeps the metallicity at which saturation is achieved rather unchanged. The resulting \Dtot--$Z$ relation in the non-linear growth regime is therefore shallower but extends over a larger range of metallicities than in the \NoDiff\ run. As discussed above, grain growth is slower in the \HiRes\ run. This might be due to differences in the driving of turbulence, which could prolong the initial dilution phase. Indeed, this phase seems to be still ongoing in this run, as indicated by dust-to-gas ratios below the stellar yield relation at relatively high $Z$ and the onset of non-linear growth at larger $Z$. The effects of non-linear growth are also visible in the halo, especially in the runs with stronger diffusion. 

At late times, in the disk, the dust abundance in the high-$Z$ particles has saturated. In the \NoDiff\ run, the shape of the \Dtot--$Z$ relation has hardly evolved compared to intermediate times, while in the runs with diffusion dilution tends to raise dust-to-metal ratio at low $Z$ above the yield relation. This is most pronounced in the \WeakDiff\ run, as in the runs with higher diffusion, there are no low-$Z$ gas particles left in the disk. In the halo, non-linear growth has left its imprint in all runs. In the \NoDiff\ run,
the relation falls just below the yield relation, with some departure towards saturation at high $Z > Z_{\sun}$. In the runs with diffusion, \Dtot reaches much lower values at low $Z$, owing to thermal sputtering efficiently lowering the dust abundance far away from the disk. The destruction is competing with the enrichment with new dust from the disk and therefore in the runs with stronger diffusion, dust abundances in the halo tend to be higher.

\subsubsection{Small-to-large grain ratio vs. metallicity}

\begin{figure}
\includegraphics[width=0.45\textwidth]{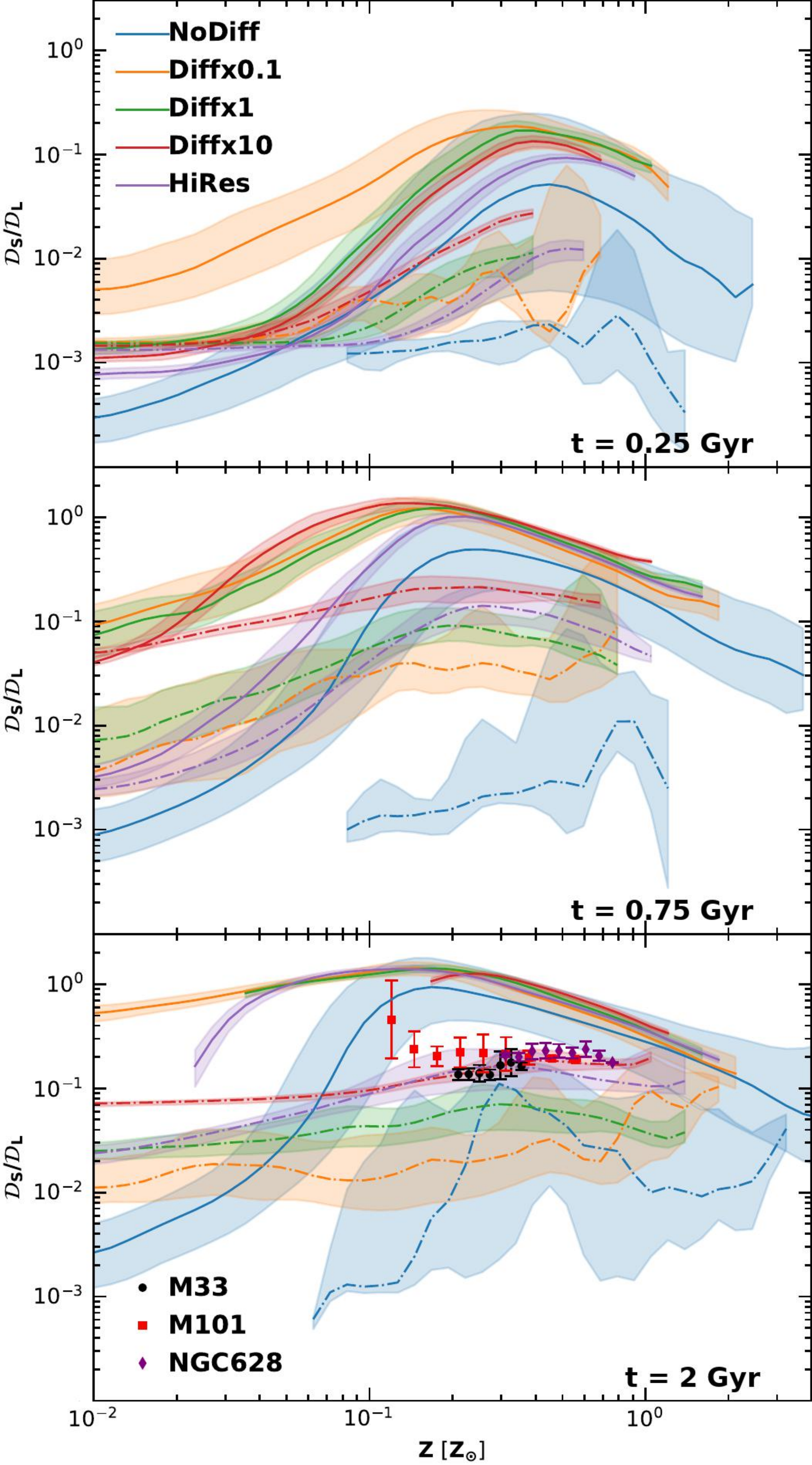}
\caption{The small-to-large grain ratio as a function of metallicity at early (top panel), intermediate (center panel) and late times (bottom panel) for the different runs. The solid lines show the median relation in the disk, while dashed lines show the relation in the halo. The shaded area corresponds to the range between the 25th and 75th percentile. Observational data shown in the bottom panel are taken from \citet{2020A&A...636A..18R} for the three galaxies as presented in the legend.}
\label{fig:DsDlvsZ}
\end{figure}

\begin{figure*}
\includegraphics[width=0.95\textwidth]{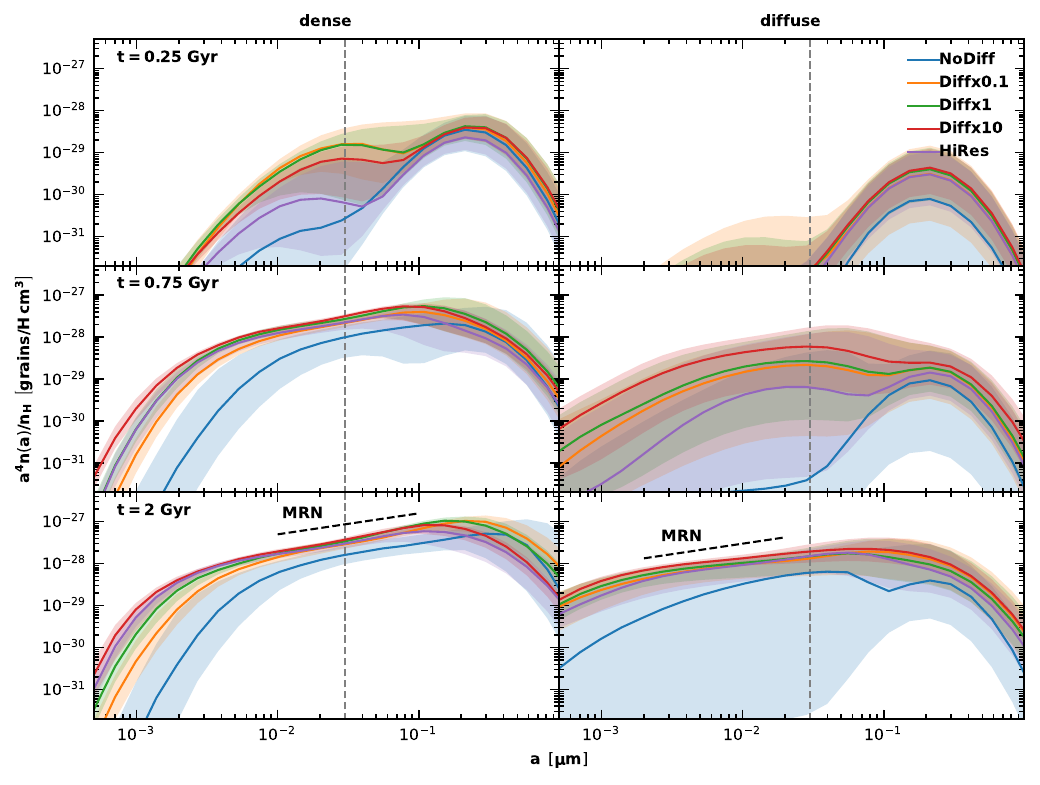}
\caption{The grain size distribution in the dense (left panels, $T < 10^3\,\text{K}\,$, $n_{\text{H}} > 10\,\text{cm}^{-3}\,$) and diffuse (right panels, $T \in \left[10^3, 10^4\right]\,\text{K}\,$, $n_{\text{H}} \in \left[0.1, 1\right]\,\text{cm}^{-3}\,$) ISM at early (top panels), intermediate (center panels) and late times (bottom panels) for the different runs. The solid lines depict the median relation. The shaded region shows the range between the 25th and 75th percentile. The black dashed line marked with `MRN' in the bottom panel shows the power-law corresponding to the \citetalias{1977ApJ...217..425M} grain-size distribution $\left[n\left(a\right) \propto a^{-3.5}\right]$. The gray dashed line indicates the division between small and large grains. }\label{fig:GSD}
\end{figure*}

The relation of the small-to-large grain ratio \DsDl with metallicity $Z$ may give hints about the evolution of the GSD. In Figure~\ref{fig:DsDlvsZ}, we show the relation between \DsDl and $Z$ in the early, intermediate and late stage of the evolution of the galaxy for each run. We show separately the relation in the disk (solid lines) and halo (dashed lines) by grouping the particles according to the criterion in Eq.~(\ref{eq:selection}). 
We further compare the relation at late times to spatially resolved data from an observational sample of late type spirals taken from \citet{2020A&A...636A..18R}. 
In the disk, the relation is generally characterised by a rise in \DsDl at low $Z < 0.1 Z_{\sun}$, while at high $Z$ it is decreased due to coagulation. This trend can be roughly seen in all runs at all times, but there are some differences as discussed below. In the halo, the relation is flatter most likely due to the importance of thermal sputtering.

At early times, in the \NoDiff\ run, \DsDl rises from $\log$\,\DsDl$\sim -4$ at low $Z$ to $\log$\,\DsDl$\sim -(1 - 2)$ at around $Z \sim 0.4 Z_{\sun}$. At higher $Z$, it then drops slightly to $\log$\,\DsDl$\sim -2$. The dispersion is rather large, spanning more than one order of magnitude. In the halo, the relation is essentially flat at $\log$\,\DsDl$\sim -3$, with some variation at larger $Z$. In the \WeakDiff\ and \StrongDiff\ runs, the relation in the disk initially shows little differences, with \DsDl being slightly higher in the \Diff\ run.
At low $Z$, the disk and halo relations are identical and flat, with constant $\log$\,\DsDl$\sim -3$ for $Z < 0.04 Z_{\sun}$. At larger metallicities, \DsDl increases and reaches its maximum value at around $Z \sim 0.3 Z_{\sun}$ before it slightly drops due to coagulation. The increase is less pronounced in the halo than in the disk, where it reaches a maximum value of $\log$\,\DsDl$\sim -1$. In the \WeakDiff\ run, the relation in the halo is similar to that in the \Diff\ run but it is slightly flatter towards large $Z$, while the relation in the disk at low $Z$ exhibits larger values, probably because it has already reached a more advanced stage in the dust growth, due to a shorter dilution phase. In the \HiRes\ run, the relation in the disk is similar to that in the \Diff\ and \StrongDiff\ runs, but it is shifted towards slightly lower values of \DsDl. The relation in the halo is similar as well, but the increase in \DsDl starts at slightly larger $Z$. Overall the inclusion of diffusion seems to slightly increase \DsDl at all metallicities, which is in line with the expectation that fluid mixing enhances dust processing. Moreover, we see an enhancement in the halo, due to a steady supply with dust grains from the disk.

At intermediate times, the relation in the halo flattens to a mild power-law in all runs with diffusion, while it develops a bump at $Z \sim Z_{\sun}$ in the \NoDiff\ run.
The power-law is shallower with stronger diffusion, but has a larger value of \DsDl$\sim 0.05 - 0.1$, whereas in the other runs with diffusion the values are lower at $\log$\,\DsDl$\sim -2$. In the \HiRes\ run, the relation in the halo has a similar shape as the one in the disk with a peak at $Z \sim 0.2 Z_{\sun}$ at $\log$\,\DsDl$\sim -1$. In the disk, the relation in the runs with diffusion at low resolution and at high resolution are very different at low $Z$, while they agree well at $Z > 0.2 Z_{\sun}$. 
In the low-resolution runs, the low-$Z$ relation is rather flat at $\log$\,\DsDl$\sim -1$ and increases towards $Z \sim 0.1 Z_{\sun}$ where it attains a value of $\log$\,\DsDl$\sim 0$, while at high resolution the value at low $Z$ is lower by almost two orders of magnitudes and increases towards a maximum of $\log$\,\DsDl$\lesssim 0$ at $Z > 0.2 Z_{\sun}$ beyond which the curves join. 
At large $Z \gtrsim 0.2 Z_{\sun}$, \DsDl falls off like \DsDl$\propto Z^{-1}$. The relation in the \NoDiff\ run
has a similar shape to the one in the \HiRes\ run, but it is overall lower by about $0.5\,\text{dex}\,$. The enhancement of \DsDl at low $Z$ and in the halo in the runs with diffusion is likely due to the diffusion of small grain abundances from high-$Z$ regions where they can grow efficiently. In the low-$Z$ regions in the disk, which are usually associated with warm diffuse gas, sputtering, which could lower the small grain abundance, is very inefficient due to low star-formation activity and temperatures that are too low for thermal sputtering. Thus small grain abundances which are comparable to the high-$Z$ growth regions can accumulate in these regions, due to a steady supply from the growth regions.

At late times, the shape of the \DsDl--Z relation in the disk in the \NoDiff\ run has hardly changed, but it has been shifted towards slightly higher \DsDl values. In the runs with diffusion, the low-$Z$ relation has flattened even further, due to the ongoing diffusion of small grains out of high-$Z$ regions, while the large-$Z$ decline with $Z^{-1}$ remains roughly the same. The range of attained metallicity values in the disk is drastically narrowed down with increasingly stronger diffusion. 
In the halo, \DsDl is almost constant with respect to $Z$ in the runs with diffusion, and takes on higher values with stronger diffusion, due to more efficient mixing with the small-grain-enriched disk. In the \NoDiff\ run, the relation in the halo is very different from the ones seen in the other runs.

None of the runs can reproduce the observations completely, which is rather surprising, given the good agreement with the radial trend of the data and the runs with diffusion shown in the bottom panel of Figure~\ref{fig:RadialDsDl}.
The slight decreasing trend of \DsDl at high metallicity in M33 and NGC 628 is consistent with coagulation, but the decrease is not as large as that predicted by the models. In the \HiRes\ and the \StrongDiff\ runs, the relation in the halo is similar to what is observed.
This might indicate that our models overestimate the effect of processes that only occur in the disk, such as accretion and coagulation. Another possible reason for the apparent failure to reproduce the observational data might be the large systematic uncertainties related to metallicity calibration. To further illustrate this point, we note, that the metallicities in NGC628 reported by \citet{2020A&A...636A..18R} and \citet[][red circles in fig. \ref{fig:RadialZ}]{2012MNRAS.423...38M} differ by an order of magnitude.

\subsubsection{Dense and Diffuse ISM}

We compare the general features of the GSD in the dense and the diffuse medium at early, intermediate, and late times among the different models. We follow the definitions of the dense and diffuse medium used by \citetalias{2020MNRAS.491.3844A}. In particular we restrict our discussion to particles within $r < 7\,\text{kpc}\,$ and $\left|z\right| < 0.3\,\text{kpc}\,$.
Particles are considered to belong to the cold, dense ISM if their density and temperature satisfy $n_\text{H} > 10 \,\text{cm}^{-3}\,$ and $T_\text{gas} < 10^3\,\text{K}\,$ and to the warm, diffuse ISM if $0.1\,\text{cm}^{-3}\, < n_\text{H} < 1 \,\text{cm}^{-3}\,$ and $10^{3}\,\text{K}\, < T_\text{gas} < 10^4\,\text{K}\,$. 

In the left panels of Figure~\ref{fig:GSD}, the GSD in the dense ISM is shown at early, intermediate, and late times. As expected, at early times the GSD is dominated by the large grains from the yield relation. In all runs, the dense GSD has a small grain tail with a bump, indicating that growth has already begun. The weight of the small grain tail differs between the runs. In the \NoDiff\ run, 
the tail hardly exceeds the yield relation, but admits a large dispersion. With diffusion, the abundance of small grains is greatly enhanced as exchange of dust between the dense and diffuse ISM enhances the processing of grains, by moving grains to where they can be processed. Without diffusion, large grains may end up trapped in dense clumps, where they may never be shattered, artificially biasing the GSD towards large grains. The enhancement is the largest with weaker diffusion, in line with the argument above, that the growth only kicks in after an initial diffusive period, where the diffusion timescale is shorter than the growth timescale. The distribution of GSD values is narrower with stronger diffusion. 

At intermediate times, dust growth has increased the abundance of small grains to a level similar to that of large grains. The overall normalisation of the GSD has increased compared to early times. At the largest grain radii, the GSD still drops with a tail similar to that of the initial log-normal distribution, but the tail towards smaller grains is now a heavy power-law tail similar to the \citetalias{1977ApJ...217..425M} grain size distribution. At the smallest grain radii, the distribution falls off significantly. The large-grain-end of the distribution shows little variation among the runs, whereas the drop at the small-grain-end occurs at larger grain radii with weaker (or no) diffusion, i.e. there are more very small grains with stronger diffusion, as exchange rates of grains between the dense and the diffuse ISM, which drive enhanced dust processing, are proportional to the diffusion strength. In the intermediate grain size range, all runs with diffusion agree remarkably well, while the GSD in the \NoDiff\ run falls short by $\sim 0.5\,\text{dex}\,$.

At late times, coagulation has kicked in, steepening the GSD and enhancing the abundance of large grains ($a > 0.1\, \micron$). This enhancement is less pronounced in the \NoDiff\ run, since the absence of mixing leads to less efficient grain processing. The power-law in the intermediate size range is closely resembling the \citetalias{1977ApJ...217..425M} grain size distribution. This is a robust prediction from theory, which shows, that collisional processes like shattering and coagulation lead to a MRN-like GSD \citep[e.g.][]{1969JGR....74.2531D, 1994Icar..107..117W, 1996Icar..123..450T, 2010Icar..206..735K}.

In the right panels of Figure~\ref{fig:GSD} the GSD in the diffuse ISM ($0.1\,\text{cm}^{-3} < n_\text{H} < 1 \,\text{cm}^{-3}\,$ and $10^{3}\,\text{K} < T_\text{gas} < 10^4\,\text{K}\,$) at early, intermediate and late times is shown. At early times, similar trends as in the case of the dense ISM are shown, though at lower normalisation. The values in the runs with diffusion are higher by almost an order of magnitude indicating that the transport of dust into the diffuse ISM is significantly more efficient with diffusion. Remarkably, while the dust enrichment of the diffuse ISM is slightly more efficient in the runs with stronger diffusion, the differences are only marginal, indicating that as long as there is even a small amount of diffusion, the diffuse ISM becomes significantly more enriched than without diffusion. 

At intermediate times, there are significant differences in the GSDs with and without diffusion. The GSD in the \NoDiff\ run is essentially the yield log-normal with a slight tail towards small grain radii. In the runs with diffusion, the yield GSD is present, but is overshadowed by a broad bump at small grain radii, centered around $a \sim 0.1 \,\micron$. With stronger diffusion, the amount of dust in the diffuse ISM is larger, because dust growth has already started in the dense ISM and grain abundances are continuously mixed throughout the ISM. This is why small grain abundances are higher in runs with stronger diffusion, where mixing is strongest. At the very small grain end, abundances of small grains are even higher than in the dense ISM, due to the absence of processes which could reduce the small grain abundance like destruction in SN shocks, which is more efficient in the dense star-forming regions. In the \HiRes\ run, the small grain bump is narrower and less pronounced than in the other runs with diffusion. This might simply be due to the later onset of dust growth which additionally is slower in this run.

At late times, coagulation kicks in, smoothing the small and large grains to form a power-law like distribution. In the \NoDiff\ run, there are two bumps; the large grain bump from the yield relation is still present, but it is slightly overshadowed by a wide small grain bump at $a \sim 0.04 \,\micron$. The abundance of small grains slowly drops towards the smallest grains, but exhibits a lot of scatter. 
Contrary to this, the GSDs in the runs with diffusion show only little scatter. In these runs, the GSD has been smoothed to a power-law that is shallower than the \citetalias{1977ApJ...217..425M} one. The reason for this shallower slope, which implies larger small-to-large-grain ratios is, that in the diffuse ISM there are no processes like sputtering due to SN shocks which could efficiently destroy small grains. At grain radii $a < 10^{-3}\,\micron$ and $a > 0.3 \,\micron$ the distribution falls off. The overall normalisation of the GSD is higher with stronger diffusion.

\section{Comparison with Observations --- Extinction curves}\label{sec:Extinction}

\begin{figure*}
\includegraphics[width=0.95\textwidth]{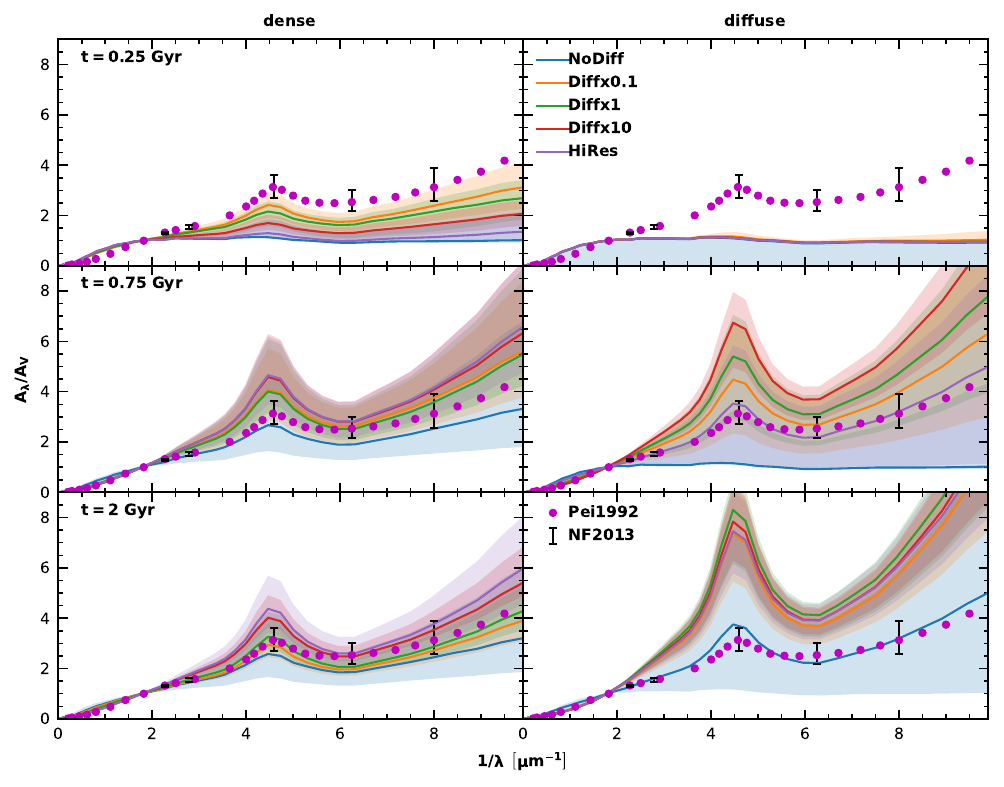}
\caption{The extinction curve in the dense (left panels, $T < 10^3\,\text{K}\,$, $n_{\text{H}} > 10\,\text{cm}^{-3}\,$) and diffuse (right panels, $T \in \left[10^3, 10^4\right]\,\text{K}\,$, $n_{\text{H}} \in \left[0.1, 1\right]\,\text{cm}^{-3}\,$) ISM at early (top panels), intermediate (center panels), and late times (bottom panels) for the different runs. The solid lines depict the median relation. The shaded region shows the range between the 25th and 75th percentile. The observed extinction curve in the Milky Way is shown by purple dots \citep{1992ApJ...395..130P}. The error bars show the dispersion along various lines of sight in the Milky Way \citep{2013ApJ...770...27N}.}\label{fig:Extinction}
\end{figure*}

We use extinction curves in order to translate the GSDs presented above into an observable that can be directly compared to observations. We compare the extinction curves obtained with the different models in the dense and diffuse medium at early, intermediate and late times to the observed extinction curve in the Milky Way. To this end, Figure~\ref{fig:Extinction} shows the extinction curves in the dense and diffuse medium. Observational data are taken from \citet{1992ApJ...395..130P} for the mean curve and from \citet{2013ApJ...770...27N} for the dispersion in the various lines of sight \citep[originally from][]{2007ApJ...663..320F}.

In the \NoDiff\ run, the extinction curve at early times is flat, both in the diffuse and in the dense medium, corresponding to the log-normal GSD from the stellar yield. With diffusion, the same is true for the extinction curve in the diffuse medium. In the dense medium however, diffusion steepens the extinction curve already at early times. This relates to the fact that the abundance of small grains of size $a \sim 0.01 \,\micron$ is already growing (Fig.~\ref{fig:GSD}), especially in the runs with weaker diffusion, where the initial dilution phase is ending earlier. 

At intermediate times, in the dense ISM the extinction curve steepens significantly. In the diffuse ISM, the extinction curve remains flat in the \NoDiff\ run, but it steepens significantly in the runs with diffusion. The steepening of the extinction curves is more pronounced in runs with stronger diffusion, corresponding to more efficient small-grain production shown in Fig.~\ref{fig:GSD}.

At late times, the extinction curves in the dense ISM flatten in the runs with diffusion, becoming comparable to the observed extinction curve in the Milky Way. This trend agrees with the findings of \citet{2021MNRAS.507..548L}, who find that the extinction curves tend to flatten at high metallicity. In the \NoDiff\ run, the extinction curve becomes slightly steeper instead, becoming slightly more comparable to the Milky Way extinction curve. The late time dense extinction curves in the \WeakDiff\ and \Diff\ runs are in best agreement with the observations around the peak and at short wavelengths, but the extinction curves in the \StrongDiff\ and \HiRes\ runs are in better agreement at long wavelengths and around $\lambda^{-1} \sim 6\, \micron^{-1}$. The median extinction curve in the \NoDiff\ run has too-low extinction at short wavelengths and too-high extinction at long wavelengths to explain the observations. This is consistent with previous results \citep[e.g.][]{2020MNRAS.491.3844A}. In the \HiRes\ run the extinction curve in the dense ISM hardly changes from intermediate to late times. While this might suggest that timescales are not converged, it needs to be stressed that the way in which the feedback model had to be adjusted in order to allow for convergence in the star formation history, might have affected the dynamics of the system in a non-trivial way, which makes it difficult to make precise statements about convergence. A detailed evaluation of the resolution dependence of the used feedback prescription is out of the scope of this work. In the diffuse ISM, the extinction curves steepen even more compared to intermediate times. In the \NoDiff\ run, the extinction curve exhibits large diversity, ranging from flat extinction curves to extremely steep ones. The median relation is comparable to the observed Milky Way extinction curve. In the runs with diffusion, the extinction curve in the diffuse ISM is extremely steep and there is only little variation among the different runs, indicating that even very weak diffusion might lead to the same result. The steepness is due to the overproduction of small grains compared with the MRN slope as shown in the right panels of Figure \ref{fig:GSD}. 

\section{Discussion and Conclusions}\label{sec:discussion}

We presented here an extension of the dust evolution model by \citetalias{2020MNRAS.491.3844A}, which aimed to address the reported underproduction of small grains and the low efficiency of coagulation. We have addressed the former by introducing a subgrid model for fluid mixing by diffusion sourced by turbulence on unresolved scales, as described in Section~\ref{sec:diffusion}. This has been motivated by the work of \citet{2021ApJ...917...12S}, who stress that SPH by design suppresses fluid mixing \citep[see also,  e.g.][]{2010MNRAS.407.1581S, 2018MNRAS.474.2194E, 2018MNRAS.480..800H}. We have addressed the latter issue by recalibrating the subgrid prescription for dense clouds to be roughly consistent with the molecular gas fraction in Milky Way mass galaxies \citep{2018MNRAS.476..875C} as described in Section~\ref{sec: MultiPhase}. As diffusion inevitably leads to the enrichment of gas in the halo with dust grains, we had to additionally extend the treatment of \citetalias{2020MNRAS.491.3844A} by destruction of dust grains by thermal sputtering \citep{1995ApJ...448...84T}. Indeed, the results of \citet{2021MNRAS.503..511G} indicate, that the omission of thermal sputtering can bias the total dust mass towards significantly higher values by promoting additional dust growth.

In order to understand the effect of diffusion, we have run a suite of simulations of an isolated Milky Way-like galaxy including a run without diffusion serving as a base line, three runs with diffusion spanning three orders of magnitude in the diffusion parameter $C_{\text{d}}$ and a high-resolution run with diffusion.
In the following we summarise our findings:
\begin{enumerate}
    \item Diffusion of metals tends to reduce stellar metallicity, while reducing the number of very metal-poor stellar populations, by reducing the width of the metallicity distribution of the (star forming) gas. This in turn leads to a slight increase in gas metallicity, bringing the metallicity of the two populations closer together to one unified value.
    \item Diffusion generally leads to narrower (i.e. less scatter) relations with less weight on the tails (i.e. less extreme values) between diffused quantities like metallicities and grain abundances.
    \item With diffusion, metals and dust extend out to larger distances from the galactic disk, and the spiral arms are traced much better by their spatial distributions, due to the reduction in particle--particle noise.
    \item Diffusion leads to the formation of a metal-poor torus in the galactic plane just beyond the edge of the gas disk. In Section~\ref{sec:profiles} we have discussed several reasons why such a structure might not have been observed so far. 
    \item Diffusion can enhance the processing of dust grains by moving grains to the sites where they can be most efficiently processed. This tends to increase small-to-large grain ratios and boosts overall dust growth, which is most efficient for smaller grains.
    \item Diffusion initially delays dust growth in an initial \emph{dilution} phase where locally high dust abundances get diluted as they spread throughout the whole disk. However, once diffusion timescales become longer than the growth timescales dilution ends, and growth by accretion starts to deplete most gas phase metals, binding them onto dust grains. Diffusion increases the fraction of gas phase metals that are depleted in this way.
    \item All of our simulations reproduce the relation between the dust-to-gas ratio and the metallicity. However, in the runs with diffusion, the metallicity  above which the dust-to-gas ratio exceeds the yield relation is lowered. This is not because growth starts at lower metallicity, but rather because of enrichment of relatively metal and dust-poor gas with dust from nearby gas with higher metallicity where growth can occur.
    \item Diffusion leads to steeper extinction curves both in the diffuse and the dense ISM. The exinction curve in the dense ISM at late times is largely consistent with the observations in the Milky Way for a value of the diffusion parameter of $C_{\text{d}} = 0.02$. However, in the diffuse ISM the extinction curve tends to be too steep, a prediction that is rather insensitive to the choice of the diffusion parameter. Observed extinction curves are measured along a line of sight, which may sample a mixture of dense and diffuse ISM. The observed extinction curves are therefore expected to lie somewhere in between our respective predictions for the dense and diffuse ISM.
    \item As indicated by the differences in the high-resolution run, the strength of diffusion does not only depend on the value of the diffusion parameter, but also on the spatial distribution of sources of metals and dust and the calibration of the feedback model driving turbulence. If the distribution of sources of metals and dust is very clumpy and metal injections tend to only affect few particles, metal gradients become large, artificially enhancing diffusion. Similarly, in the presence of strong kinetic feedback, driving strong turbulence, diffusion can become significantly enhanced. The exact value of the diffusion parameter therefore needs to be calibrated along with the feedback model. 
    \item In all of our models, the radial profiles of dust and metal abundances as well as the small-to-large grain ratios are in reasonable agreement with the observations in nearby galaxies from \citet{2012MNRAS.423...38M} and \citet{2020A&A...636A..18R}. The dust-to-metal ratios in our models tend to be slightly larger than the observed values taken from \citet{2012MNRAS.423...38M}. Curiously, from Figure~\ref{fig:DsDlvsZ} we can see that in our simulations the relation between \DsDl and $Z$ in the disk tends to be above the observational relation \citep{2020A&A...636A..18R}, while the relation in the halo lies below. While this might indicate that in the observations some kind of average between the two relations is taken, which would explain the slightly lower values, it seems more likely that a better calibration of our subgrid model, a better treatment of dust processing due to SN feedback and the inclusion of AGN feedback, will lead to a better agreement with the observations.
\end{enumerate}
In this work, we employed an isotropic diffusion model. While this might be appropriate for gas phase metals and small grains which are closely coupled to the gas, large dust grains can decouple from small-scale turbulence eddies
\citep[e.g.][]{2015MNRAS.449.1625B, 2015MNRAS.452.3932B, 2018MNRAS.478.2851M}. It might thus be worthwhile to model the effects of drag force on the dust grains as an anisotropic diffusion current. Moreover, throughout this paper we have assumed neutral grains and neglected grain charging. \citet{2022MNRAS.510.1068G} have found that dust grains in \ion{H}{ii} regions can pick up considerable amounts of negative charges. The charging of grains can have a profound impact on interaction rates, which might affect both grain growth as well as dust-based heating- and cooling-rates. Therefore, it might be interesting to include the effects of grain charing into future dust evolution models. Finally, in this work we assumed that dust grains were compact spheres. \citet{2022MNRAS.509.5771H} developed a model for the grain-size-dependent evolution of grain porosity. They show that porosity may act to steepen the extinction curve.
Grain porosity might also have important implications for other processes like molecule formation on grain surfaces through the change of grain surface area. It might thus be advisable to take the effect of porosity into account in future studies.

In conclusion, including diffusion of dust and metals in our simulations lead to an enhancement in the production of small grains, while being largely consistent with total dust and metal abundances in previous works \citep[e.g.][]{2017MNRAS.466..105A, 2019MNRAS.485.1727H, 2020MNRAS.491.3844A}. There are some discrepancies between our simulations and the observations. In particular, we have found that in the outer parts of the galactic disk, the dust-to-metal ratios and the small-to-large grain ratios tend to be too high in our simulations, indicating that production of small grains is too efficient in this regime. Future analysis will show how this issue can be resolved. It should also be stressed that, while the setting of an isolated disk galaxy is a useful framework for testing new models, its unrealistic initial conditions and the absence of environmental effects like mergers, filaments as well as large-scale tidal fields limit its applicability to realistic comparisons with observations. In order to obtain results that allow for a reasonable comparison with nearby spirals, future cosmological zoom-in simulations will have to be performed. Despite these limitations, we are confident that our results regarding the effects of diffusion hold true and hope that they will be useful to guide future efforts for modelling the evolution of dust and the GSD in the ISM.

\section*{Acknowledgements}

We thank the referee for his insightful comments and suggestions that helped to improve the quality of this communication. We acknowledge Shohei Aoyama and Ikkoh Shimizu who provided their versions of {\sc GADGET3-Osaka}, which served as useful references for the implementation of the feedback and dust evolution model in our {\sc GADGET4-Osaka}.
Our numerical simulations and analyses were carried out on our local cluster {\sc Orion}.
This work was partly supported by the JSPS KAKENHI Grant Number JP17H01111, 19H05810, 20H00180. 
KN acknowledges the support from the Kavli IPMU, World Premier Research Center Initiative (WPI).
HH thanks the Ministry of Science and Technology (MOST) for support through grant
MOST 108-2112-M-001-007-MY3, and the Academia Sinica
for Investigator Award AS-IA-109-M02.

\section*{Data Availability}

Data related to this publication and its figures are available on request from
the corresponding author.



\bibliographystyle{mnras}
\bibliography{references_v1} 

\clearpage
\appendix

%
\section{Thermal Sputtering}\label{Appendix:thermsputtering}
We express the grain mass distribution in terms of the mass density associated to the $i$th grain size bin $\mathcal{M}_i\left(t\right) = \rho_{d}\left(m_i, t\right) m_i \Delta\mu$ with $\mu = \text{ln}\, m$. 
Discretizing equation~(\ref{eq:sputtering}) for $\mathcal{M}_i$ in $\mu$ yields the linear ODE:
\begin{equation}\label{eq:discrSput}
    \frac{\text{d}\mathcal{M}_i\left(\Tilde{t}\right)}{\text{d} \Tilde{t}} =  \mathcal{M}_{i+1}\left(\Tilde{t}\right) \gamma_{i+1} \delta - \mathcal{M}_{i}\left(\Tilde{t}\right)\gamma_{i}.
\end{equation}
Here $\Tilde{t} = t / \Delta\mu$, $\delta = m_{i+1}/m_{i} = {\rm const}.$ and $\gamma_i = \tau^{-1}\left(m_i\right)$. 
From equation (\ref{eq:discrSput}) it is clear that $\mathcal{M}_i$ only depends on the initial conditions of $\mathcal{M}_j$ with $j \geq i$ and in particular $\mathcal{M}_N\left(\Tilde{t}\right) / \mathcal{M}_{N, 0} = e^{-\gamma_N \Tilde{t}}$. Thus the ODE can be solved with the ansatz
\begin{equation}
    \mathcal{M}_i = \sum_{k = i}^{N} G_{ik} f_k\left(\Tilde{t}\right).
\end{equation}
Here $f_i\left(\Tilde{t}\right) = f_{i, 0} e^{-\gamma_i \Tilde{t}}$ incorporates the initial conditions and decay in each grain size bin, and
\begin{equation}
    G_{ik} = \delta^{k - i} \prod_{j = 0}^{k-i-1} \frac{\gamma_{k - j}}{\gamma_{i + j} - \gamma_k}
\end{equation}
determines how much of the $k$th bin gets deposited in the $i$th bin.

\section{Particle Selection}\label{Appendix:Selection}
\begin{figure}
\includegraphics[width=0.45\textwidth]{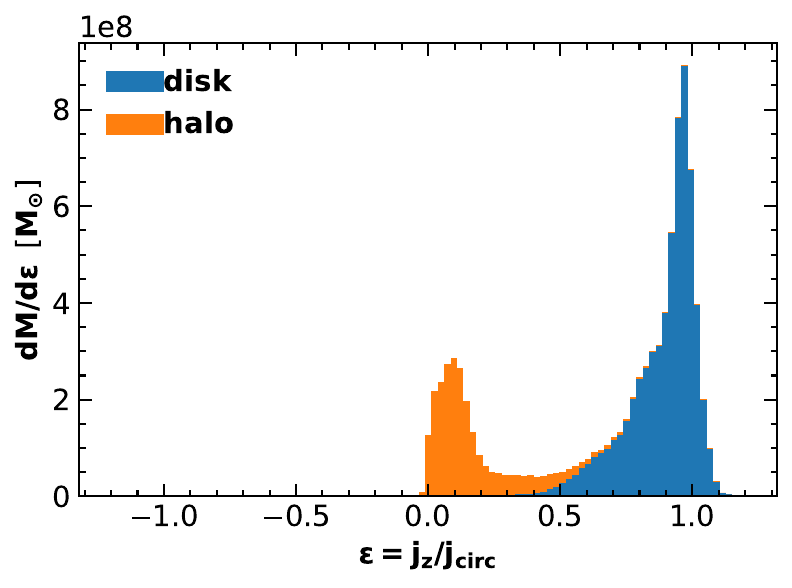}
\caption{
The distribution of orbital circularity parameter of the gas particles in the disk (blue) and the halo (orange). Histograms are stacked on top of each other. The disk component is selected according to the thermodynamic criterion described in Section~\ref{sec:Selection}, the halo is the complement.}\label{fig:Circularity}
\end{figure}
\begin{figure}
\includegraphics[width=0.45\textwidth]{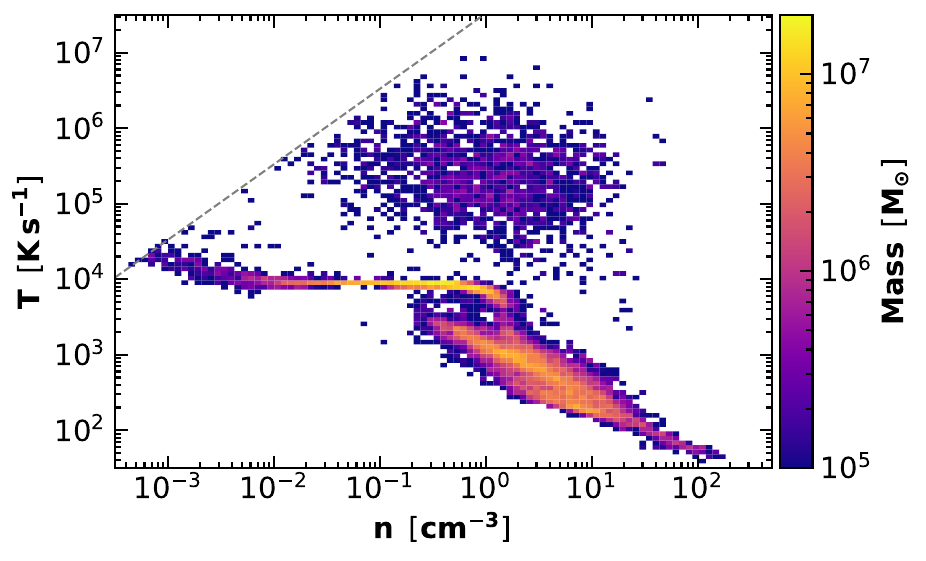}
\caption{
Phase diagram of gas belonging to the disk according to the thermodynamic selection criterion (described in Sec~\ref{sec:Selection}) and with $\epsilon < 0.8$. Cells are color coded according to the total mass of particles in each cell. The gray line indicates the threshold for the thermodynamic cut. The gas shown in this figure will be missed if only $\epsilon > 0.8$ is used for the particle selection. }\label{fig:EoS}
\end{figure}
In Section~\ref{sec:Selection} we have described a criterion to select gas particles belonging to the galactic disk based on the equation of state of the gas. In the literature, a commonly used kinematic selection criterion for the \emph{thin} disk component of galaxies is based on cuts on the orbital circularity parameter $\epsilon = j_z / j_{\text{circ}}\left(E\right)$. The idea behind the kinematic criterion is that particles belonging to the thin disk are rotating around the galactic spin axis with a specific angular momentum close to the one corresponding to a circular orbit with the same gravitational binding energy, whereas other components like a spheroid or a bar would have on average a much lower specific angular momentum. Thus selecting particles with an orbital circularity of say $\epsilon > 0.8$ will result in a quite pure selection of the thin disk component. In the present study, we are interested in all components of the disk including any potential spheroidal or bar components, but aim to exclude the contributions from the halo. While the criterion based on orbit circularity is surely expected to exclude any contribution from the halo, it might also exclude other components of the galactic disk. In this appendix we compare the two criteria and motivate, why we have chosen the former criterion over the latter.

\begin{figure}
\includegraphics[width=0.45\textwidth]{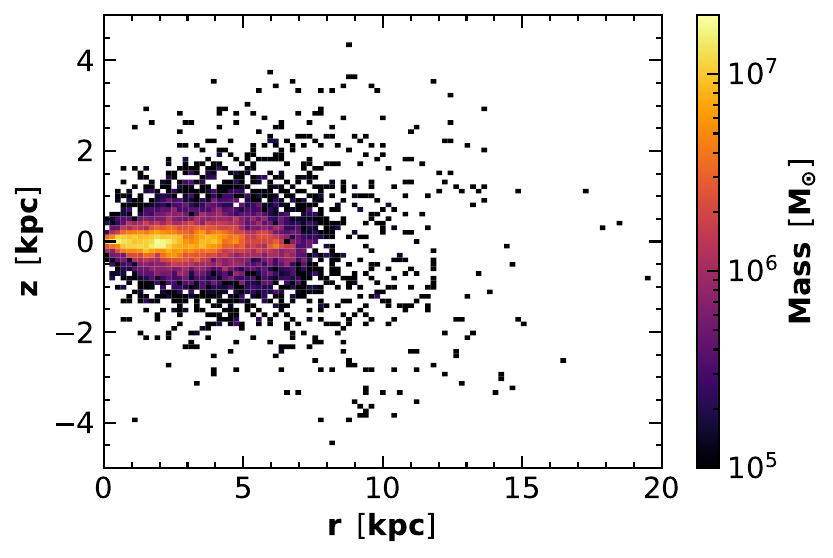}
\caption{
The spatial distribution of gas particles belonging to the disk according to the thermodynamic selection criterion and with $\epsilon < 0.8$ in the $r$--$z$ plane. The color coding indicates the total mass of the particles belonging to each bin. }\label{fig:r-z-plane}
\end{figure}

In order to further explore this, Figure~\ref{fig:Circularity} shows the circularity histogram of the disk and halo components as selected by the thermodynamic selection criterion. The disk component is peaked at $\epsilon = 1$, consistent with the kinematic selection criterion, but extends down to $\epsilon \sim 0.5$, with roughly 20 percent of its mass below $\epsilon = 0.8$. The halo is peaked just above $\epsilon = 0$, with negligible contributions above $\epsilon = 0.7$. Thus, we can conclude that the kinematical criterion is more stringent than the thermodynamic one. 
A phase diagram of the gas, selected with our thermodynamic selection criterion, that would be excluded by the kinematic one ($\epsilon < 0.8$) is shown in Figure~\ref{fig:EoS}. As can be seen, the kinematic criterion would exclude large amounts of gas with temperatures well below $10^4\,\text{K}$, which are expected to be part of the galactic disk. In order to verify this, in Figure~\ref{fig:r-z-plane} we show the spatial distribution of the particles which satisfy the thermodynamic criterion, but not the circularity criterion $\epsilon > 0.8$. We find that most particles selected in this way are located in a thin disk with radius $r \sim 5\,\text{kpc}\,$. The dynamics of these particles might be more strongly influenced by the central bulge which might affect the circularity of their orbits.

Therefore, we can conclude, that for the purposes of analyzing the gas within the galactic disk, the thermodynamic selection criterion is more reliable than the more strict kinematic one.

\section{Metal-poor  Torus}\label{Appendix:LowZTorus}

\begin{figure}
\includegraphics[width=0.45\textwidth]{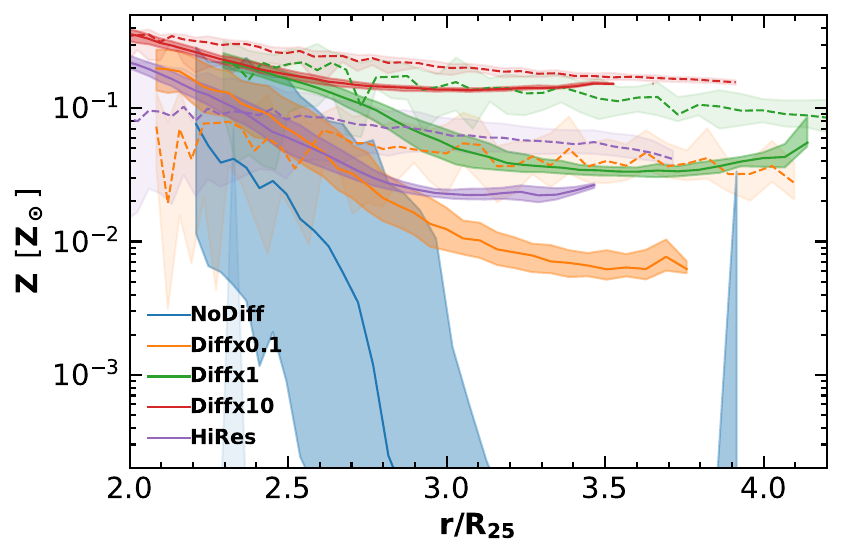}
\caption{
The radial metallicity profiles of each run, zoomed in around the location of the metal-poor torus. We show separately the profile in the disk (solid lines) and halo (dashed lines) as selected by the thermodynamic selection criterion. The shaded area depicts the range of values between the 25th and 75th percentile.}\label{fig:torusTH}
\end{figure}

\begin{figure}
\includegraphics[width=0.45\textwidth]{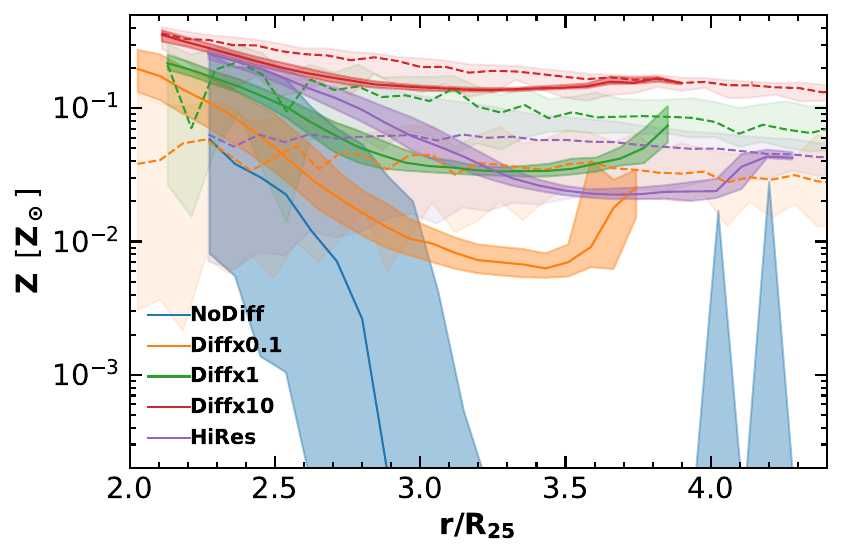}
\caption{
Same as Fig.~\ref{fig:torusTH} but with components selected with the kinematic selection criterion.}\label{fig:torusKIN}
\end{figure}

In Section~\ref{sec:SpatialDistribution} we have shown that, in the runs with diffusion, a toroidal region with low metallicity that we called a metal-poor torus forms around the edges of the disk. We have already seen in Figure~\ref{fig:ShearMap} that in this region the shear is particularly weak and thus we expect only little mixing, while at larger radii and above and below the disk there is more shear and therefore the gas will be better mixed. To verify that this is indeed a feature of the disk and not an effect due to the superposition of the well-mixed halo gas and the unmixed disk, in this appendix we show the metallicity profiles of the disk and halo component, respectively. We select each component with the thermodynamic selection criterion, and in order to make sure that the result is independent of the employed selection criterion, also with the kinematic selection criterion.

In Figure~\ref{fig:torusTH} we show separately the radial metallicity profile of the disk and halo component selected by utilizing the thermodynamic selection criterion. We have zoomed into the region where the metal-poor torus is located. The gas in the halo is well mixed and has an almost flat profile, with metallicities around $Z_\text{halo} \sim 0.1 Z_{\sun}$. The profile in the disk is characterised by a steady decline towards larger radii, with a minimum around $R_\text{min} \sim \left(3 - 4\right) R_{25}$. The minimum value of the metallicity falls below the value of the metallicity in the halo and tends to be lower with weaker diffusion (with the extreme case of $Z_\text{min} = 10^{-4} Z_{\sun}$, the adopted value of the metallicity floor in the \NoDiff\ run). Beyond this minimum in all runs, the metallicity increases, due to mixing with the halo, which is in line with Figure~\ref{fig:ShearMap}, which shows that the shear increases beyond $R \sim 15 \,\text{kpc}\,$.

Figure~\ref{fig:torusKIN} shows the same plot as Figure~\ref{fig:torusTH}, but here the disk and halo component are selected by utilizing the kinematic selection criterion. There are only minor differences between the two, indicating, that our conclusion that the metal-poor torus is indeed corresponding to an unmixed region \emph{within the disk} where there is only little shear is robust with respect to the particle selection criterion.

\bsp	
\label{lastpage}
\end{document}